\documentclass[a4paper]{iopart}

\usepackage{amssymb}
\usepackage{latexsym}
\usepackage{graphicx}
\usepackage{times}

\begin{document}


\title{On the gravitational radiation from the collapse of neutron
  stars to rotating black holes}


\author{Luca~Baiotti} 
\address{Max-Planck-Institut f\"ur Gravitationsphysik,
Albert-Einstein-Institut, 14476 Golm, Germany}

\author{Ian~Hawke}
\address{School of Mathematics, University of Southampton,
Southampton SO17 1BJ, UK}

\author{Luciano~Rezzolla}
\address{Max-Planck-Institut f\"ur Gravitationsphysik,
Albert-Einstein-Institut, 14476 Golm, Germany}
\address{Department of Physics, Louisiana State University, Baton
Rouge, LA 70803 USA}



\date{\today}


\begin{abstract}
  We provide details and present additional results on the numerical
  study of the gravitational-wave emission from the collapse of
  neutron stars to rotating black holes in three dimensions. More
  specifically, we concentrate on the advantages and disadvantages of
  the use of the excision technique and on how alternative approaches to
  that of excision can be successfully employed. Furthermore, as a
  first step towards source-characterization, we present a systematic
  discussion of the influence that rotation and different
  perturbations have on the waveforms and hence on the energy emitted
  in gravitational waves.
\end{abstract}


\section{Introduction}
\label{sec:introduction}

        The study of the gravitational collapse of rotating stars to
black holes is a cornerstone of any theory of gravity and a long standing
problem in General Relativity. Important issues in relativistic
astrophysics awaiting clarification, such as the mechanism responsible
for $\gamma$-ray bursts, may be unveiled with a more detailed
understanding of the physics of gravitational collapse in rotating and
magnetized stars. Furthermore, the study of gravitational collapse will
provide the waveforms and the energetics of one of the most important
sources of gravitational radiation.

In our previous work~\cite{Baiotti04b,Baiotti06}, we have described how
we can perform accurate three-dimensional relativistic simulations of
such events and how we are able to extract their gravitational wave
signals. Before our work, the only work in the literature about the
gravitational radiation from neutron-star collapse dates back 20 years
and is restricted to axisymmetry~\cite{Stark85}. Here, after a brief
introduction to our code and to the models we have simulated, we give
more details on the techniques and results presented
in~\cite{Baiotti04b,Baiotti06}, focussing on the properties of the
gravitational waves produced and on how these are influenced by factors
such as rate of rotation of the compact star or the type and amplitude of
the perturbations introduced to trigger the collapse.

Throughout the paper we use a spacelike signature $(-,+,+,+)$ and a
system of units in which $c=G=M_\odot=1$ (unless explicitly shown
otherwise for convenience). Greek indices are taken to run from 0 to 3,
Latin indices from 1 to 3 and we adopt the standard convention for the
summation over repeated indices.


\section{Basic equations and their implementation}
\label{equations}
\medskip

        The {\tt Whisky} code solves the general relativistic
hydrodynamics equations on a 3D numerical grid with Cartesian
coordinates~\cite{Baiotti03a}. The code has been constructed within the
framework of the {\tt Cactus} Computational Toolkit (see
refs.~\cite{Goodale02a, cactusweb1} for details), and it is developed at
the Albert Einstein Institute and at the Louisiana State University. This
public domain code provides high-level facilities such as parallelization,
input/output, portability on different platforms and several evolution
schemes to solve general systems of partial differential
equations. Clearly, special attention is dedicated to the solution of the
Einstein equations, whose matter-terms in non-vacuum spacetimes are
handled by the {\tt Whisky} code. 

In other words, while  the {\tt Cactus} code provides at each time step
and on a spatial hypersurface the solution of the Einstein equations
\begin{equation}
\label{efes}
G_{\mu \nu}=8\pi T_{\mu \nu}\ , 
\end{equation}
where $G_{\mu \nu}$ is the Einstein tensor and $T_{\mu \nu}$ is the
stress-energy tensor, the {\tt Whisky} code provides the time evolution
of the hydrodynamics equations, expressed through the conservation
equations for the stress-energy tensor $T^{\mu\nu}$ and for the matter
current density $J^\mu$
\begin{equation}
\label{hydro eqs}
\nabla_\mu T^{\mu\nu} = 0\;,\;\;\;\;\;\;\;\;\;\;\;\;
\nabla_\mu J^\mu = 0\;.
\end{equation}

        In what follows, and mostly for the sake of completeness, we give
a brief overview of how both the right and the left-hand-side of
equations (\ref{efes}) are computed within the coupled {\tt
Cactus/Whisky} codes. The equations presented have already been discussed
in several different publications, e.g. in
~\cite{Alcubierre99d,Baiotti04,Baiotti04b} and we refer the interested
readers to these works for more details.

\subsection{Evolution of the field equations}   
\label{feqs}

Many different formulations of the equations have been proposed
throughout the years, starting with the ADM formulation in
1962~\cite{Arnowitt62}. As mentioned in the Introduction, we use the NOK
\cite{Nakamura87} formulation, which is based on the ADM construction and
has been further developed in~\cite{Shibata95}.

Details of our particular implementation of the conformal traceless
reformulation of the ADM system as proposed
by~\cite{Nakamura87,Shibata95,Baumgarte99} are extensively described
in~\cite{Alcubierre99d,Alcubierre02a} and will not be repeated here. We
only mention, however, that this formulation makes use of a conformal
decomposition of the three-metric, \hbox{$\tilde \gamma_{ij} = e^{- 4
\phi} \gamma_{ij}$}, and the trace-free part of the extrinsic curvature,
\hbox{$A_{ij} = K_{ij} - \gamma_{ij} K/3$}, with the conformal factor
$\phi$ chosen to satisfy $e^{4 \phi} = \gamma^{1/3}$, where $\gamma$ is
the determinant of the spatial three-metric $\gamma_{ij}$. In this
formulation, in addition to the evolution equations for the conformal
three-metric $\tilde \gamma_{ij}$ and the conformal traceless extrinsic
curvature $\tilde A_{ij}$, there are evolution equations for the
conformal factor $\phi$, for the trace of the extrinsic curvature $K$ and
for the ``conformal connection functions'' $\tilde \Gamma^i \equiv \tilde
\gamma^{ij}{}_{,j}$. We note that although the final mixed, first and
second-order, evolution system for the variables $\left\{ \phi, K, \tilde
\gamma_{ij}, {\tilde A_{ij}}, {\tilde \Gamma^i} \right\}$ is not in any
immediate sense hyperbolic, there is evidence showing that the
formulation is at least equivalent to a hyperbolic
system~\cite{Sarbach02a,Bona:2003qn,Nagy:2004td}. In the formulation
of~\cite{Shibata95}, the auxiliary variables ${\tilde F}_i = -\sum_j
{\tilde \gamma_{ij,j}}$ were used instead of the ${\tilde \Gamma^i}$.

\subsubsection{Gauge choices}

        The code is designed to handle arbitrary shift and lapse
conditions, which can be chosen as appropriate for a given spacetime
simulation.  More information about the possible families of spacetime
slicings which have been tested and used with the present code can be
found in~\cite{Alcubierre99d,Alcubierre01a}. Here, we limit ourselves to
recalling details about the specific foliations used in the present
evolutions. In particular, we have used hyperbolic $K$-driver slicing
conditions of the form
\begin{equation}
\partial_t \alpha = - f(\alpha) \;
\alpha^2 (K-K_0),
\label{eq:BMslicing}
\end{equation}
with $f(\alpha)>0$ and $K_0 \equiv K(t=0)$. This is a generalization of
many well known slicing conditions.  For example, setting $f=1$ we
recover the ``harmonic'' slicing condition, while, by setting
\mbox{$f=q/\alpha$}, with $q$ an integer, we recover the generalized
``$1+$log'' slicing condition~\cite{Bona94b}.  In particular, all of the
simulations discussed in this paper are done using condition
(\ref{eq:BMslicing}) with $f=2/\alpha$. This choice has been made mostly
because of its computational efficiency, but we are aware that ``gauge
pathologies'' could develop with the ``$1+$log''
slicings~\cite{Alcubierre97a,Alcubierre97b}.


        For the spatial gauge, we use one of the ``Gamma-driver''
shift conditions proposed in~\cite{Alcubierre01a} (see also 
\cite{Alcubierre02a}), that essentially
act so as to drive the $\tilde{\Gamma}^{i}$ to be constant. In this
respect, the ``Gamma-driver'' shift conditions are similar to the
``Gamma-freezing'' condition $\partial_t \tilde\Gamma^k=0$, which, in
turn, is closely related to the well-known minimal distortion shift
condition~\cite{Smarr78b}. The differences between these two conditions
involve the Christoffel symbols and are basically due to the fact that the
minimal distortion condition is covariant, while the Gamma-freezing
condition is not.

        In particular, all the results reported here have been
obtained using the hyperbolic Gamma-driver condition,
\begin{equation}
\partial^2_t \beta^i = F \, \partial_t \tilde\Gamma^i - \eta \,
\partial_t \beta^i,
\label{eq:hyperbolicGammadriver}
\end{equation}
where $F$ and $\eta$ are, in general, positive functions of space and
time. For the hyperbolic Gamma-driver conditions it is crucial to add a
dissipation term with coefficient $\eta$ to avoid strong oscillations in
the shift. Experience has shown that by tuning the value of this
dissipation coefficient it is possible to almost freeze the evolution of
the system at late times. We typically choose $F={3}/{4}$ and $\eta=3$
and do not vary them in time.

The singularity-avoiding properties of the above gauge choices have
proved equally good both when using excision, as we did in
refs.~\cite{Baiotti04} and~\cite{Baiotti04b}, and when not using
excision. In this latter case, the addition of a small amount of
dissipation in the metric and gauge terms is necessary to obtain
long-term stable evolutions~\cite{Baiotti06}. In the absence of an
excised region of spacetime, the gauge
choices~(\ref{eq:hyperbolicGammadriver}) are essential to ``freeze'' the
evolution in those regions of the computational domain inside the
apparent horizon, where the metric functions experience the growth of
very large gradients.

\subsection{Evolution of the hydrodynamics equations}    

        An important feature of the {\tt Whisky} code is the
implementation of a \textit{conservative formulation} of the
hydrodynamics equations \cite{Marti91,Banyuls97,Ibanez01}, in which the
set of equations (\ref{hydro eqs}) is written in a hyperbolic,
first-order and flux-conservative form of the type
\begin{equation}
\label{eq:consform1}
\partial_t {\mathbf q} + 
        \partial_i {\mathbf f}^{(i)} ({\mathbf q}) = 
        {\mathbf s} ({\mathbf q})\ ,
\end{equation}
where ${\mathbf f}^{(i)} ({\mathbf q})$ and ${\mathbf s}({\mathbf q})$
are the flux-vectors and source terms, respectively~\cite{Font03}.  Note
that the right-hand-side (the source terms) depends only on the metric,
and its first derivatives, and on the stress-energy tensor. Furthermore,
while the system (\ref{eq:consform1}) is not strictly hyperbolic,
strong hyperbolicity is recovered in a flat spacetime, where ${\mathbf s}
({\mathbf q})=0$.

        Additional details of the formulation we use for the
hydrodynamics equations can be found in ref.~\cite{Font03}. We stress
that an important feature of this formulation is that it allows for the
extension to a general relativistic context the powerful numerical methods
developed in classical hydrodynamics, in particular High-Resolution
Shock-Capturing schemes based on
exact~\cite{Marti99,Rezzolla01,Rezzolla03} or approximate Riemann solvers
(see ref.~\cite{Font03} for a detailed bibliography). Such schemes are
essential for a correct representation of shocks, whose presence is
expected in several astrophysical scenarios.

For all the results presented here, we have solved the hydrodynamics
equations employing the Marquina flux formula and a third-order
PPM~\cite{Colella84} reconstruction, and the Einstein field equations
using a Runge-Kutta scheme of third order, the ``$1+$log'' slicing
condition and the ``Gamma-driver'' shift
conditions~\cite{Alcubierre01a}. After having seen no significant
difference in the dynamics of our models while using polytropic or
ideal-fluid EOSs (because no shocks form), we have concentrated only
on the former, which require slightly smaller computational times.

\subsection{Mesh Refinement}    
\label{sec:mesh_refinement}

An important improvement with respect to the work
presented~\cite{Baiotti04}, which we refer to as paper I hereafter, is
the possibility of solving now both the fields and hydrodynamics
equations on non-uniform grids using a ``box-in-box'' mesh refinement
strategy~\cite{Schnetter-etal-03b} (see Fig.~2 of~\cite{Baiotti06c}).
All of the simulations of paper I were redone with the new grid setup and
no differences with respect to the unigrid results were found as far as
the dynamics of the matter and of the horizons are concerned. On the
other hand, this change introduces two important advantages: firstly, it
reduces the influence of inaccurate boundary conditions at the outer
boundaries which can be moved far from the central source; secondly, it
allows for the wave zone to be included in the computational domain and
thus for the extraction of important information about the gravitational
wave emission produced during the collapse.

In practice, we have adopted a Berger-Oliger prescription for the
refinement of meshes on different levels~\cite{Berger84} and used the
numerical infrastructure described in~\cite{Schnetter-etal-03b}, i.e.,
the {\tt Carpet} mesh refinement driver for {\tt Cactus} (see
\cite{carpetweb} for details). In addition to this, in
ref.~\cite{Baiotti04b} we had also used a simplified form of
adaptivity in which new refined levels are added at predefined
positions during the evolution.  This progressive mesh refinement,
which allows to use much less computational resources, was the key
improvement to our previous code~\cite{Baiotti04} and allowed to
extract, for the first time in 3D calculations, the gravitational
waveform from the collapse to a rotating black hole. While a fixed or
a progressive mesh-refinement technique leads to no appreciable
change in the dynamics of the matter or of the horizons, it can
influence the spectral distribution of the radiation emitted
especially at high frequencies~\cite{Baiotti06c}.  All of the results
presented here have been computed using seven fixed levels of
refinements.

\subsection{Singularity Excision}    
\label{sec:exc+dissip}

The use of the excision technique was essential in paper I for studying
the dynamics of the collapse with uniform grids, because these, combined
with the computational resources available at that time, had forced us to
use outer boundaries close to the stellar surface and a relatively coarse
resolution. Such a resolution was sufficient to describe accurately the
dynamics of the matter and of the horizons but also required the use of
excision if the simulation was to be carried out beyond horizon
formation.

An alternative to the use of the excision technique consists in adding a
small amount of dissipation to the evolution equations for the metric and
gauge variables and of relying on the use of singularity-avoiding gauges
and of high resolution to extend the simulations well past the formation
of the apparent horizon. More specifically, we have used an artificial
dissipation of the Kreiss--Oliger type~\cite{Kreiss73} on the
right-hand-sides of the evolution equations for the spacetime variables and the
gauge quantities. This is needed mostly because all the field variables
develop very steep gradients in the region inside the apparent
horizon. Under these conditions, small high-frequency oscillations
(either produced by finite-differencing errors or by small reflections
across the refinement or outer boundaries) can easily be amplified, leave
the region inside the apparent horizon and rapidly destroy the
solution. In practice, for any time-evolved quantity $u$, the
right-hand-side of the corresponding evolution equation is modified with
the introduction of a term of the type ${\cal L}_{\mbox{\tiny diss}}(u) =
-\varepsilon h^3 \partial^4_i u$, where $h$ is the grid spacing, and
$\varepsilon$ is the dissipation coefficient, which is allowed to vary in
space.

\begin{figure}
\centering
\includegraphics[angle=0,width=8.5cm]{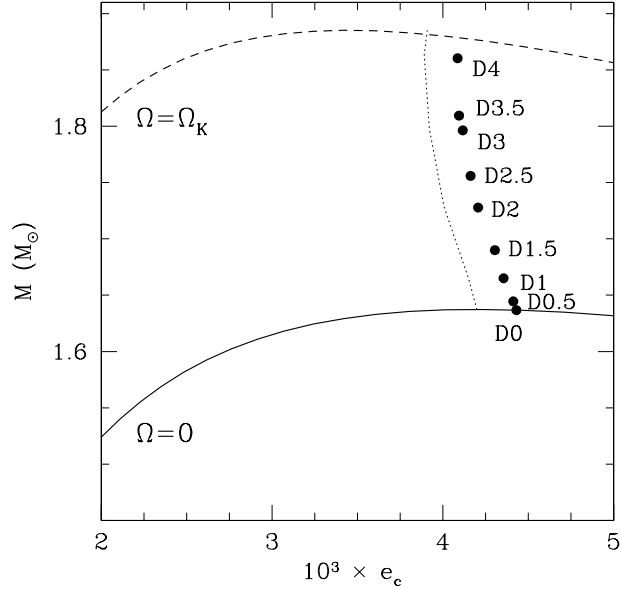} 
\caption{Gravitational mass shown as a function of the central energy
       density for equilibrium models constructed with the polytropic
       EOS, for $\Gamma=2$ and polytropic constant $K_{\rm ID}=100$. The
       solid, dashed and dotted lines correspond to the sequence of
       non-rotating models, the sequence of models rotating at the
       mass-shedding limit and the sequence of models that are at the
       onset of the secular instability to axisymmetric
       perturbations. Also shown are the dynamically unstable (filled
       circles) initial models used in the collapse simulations.}
\label{fig:Initial} 
\end{figure}

We have experimented with configurations in which the coefficient was
either constant over the whole domain or larger for the grid points
inside the apparent horizon. We noticed no significant difference between
these two cases. Much more sensitive is instead the choice of the
\emph{value} of $\varepsilon$. In the simulations reported here, the
employed values of $\varepsilon$ are between 0.0075 and 0.02. For each
initial model, two values of $\varepsilon$, $\varepsilon_{\textrm{min}}$
and $\varepsilon_{\textrm{max}}$, can be determined, such that for values
smaller than $\varepsilon_{\textrm{min}}$ the dissipation is not strong
enough to cure the instability, and such that for values larger than
$\varepsilon_{\textrm{max}}$ the solution is different from the one
obtained without dissipation (overdissipation). Such differences in the
solution cannot be seen in the dynamics of the matter or of the horizons,
but only in the very sensitive waveforms.

The use of numerical dissipation stops the growth of the metric
functions, which, instead of growing more and more while approaching
the singularity, stabilise to a stationary state. Outside the horizon,
the spacetime is practically identical to the one obtained without
dissipation and the dynamics of the horizon itself is the same as in
the case in which excision was performed (up to when the latter data
are available). On the contrary, the metric inside the horizon is
rather far from being a solution of the Einstein equations, but this
does not influence the outside spacetime, as shown also in Fig.~2 of
ref.~\cite{Baiotti06}. Hereafter, all of the presented results will
refer to simulations carried out without excision and we note that no
dissipation is added to the evolution of any matter variable.

\begin{table}
\begin{center}
\caption{Equilibrium properties of the initial stellar models. The
        different columns refer respectively to: the central rest-mass
        density $\rho_c$, the ratio of the polar to equatorial coordinate
        radii $r_p/r_e$, the gravitational mass $M$, the circumferential
        equatorial radius $R_e$, the angular velocity $\Omega$, the ratio
        $J/M^2$ where $J$ is the angular momentum, the ratio of
        rotational kinetic energy to gravitational binding energy
        $T/|W|$. All models have been computed with a polytropic EOS with
        $K_{\rm ID}=100$ and $\Gamma=2$.}
\bigskip
\begin{tabular}{{l}*{7}{c}}
\hline
Model & $\rho_c(\times 10^{-3})$   & $r_p/r_e$ & $M$ & $R_e$ &
        $\Omega(\times 10^{-3})$  &$J/M^2$
        & $T/|W|(\times 10^{-2})$ \\ 
\hline 
\hline 
$D0$   & 3.325 & 1.00 & 1.636 & 7.54 & 0.00 & 0.000 & 0.00 \\
$D0.5$ & 3.314 & 0.99 & 1.644 & 7.59 & 0.92 & 0.108 & 0.32 \\
$D1$   & 3.280 & 0.95 & 1.665 & 7.74 & 1.73 & 0.206 & 1.16 \\
$D1.5$ & 3.249 & 0.91 & 1.690 & 7.91 & 5.76 & 0.281 & 2.13 \\
$D2$   & 3.189 & 0.85 & 1.728 & 8.21 & 2.88 & 0.362 & 3.52 \\
$D2.5$ & 3.162 & 0.81 & 1.756 & 8.43 & 3.20 & 0.410 & 4.48 \\
$D3$   & 3.134 & 0.75 & 1.797 & 8.80 & 3.55 & 0.468 & 5.79 \\
$D3.5$ & 3.121 & 0.73 & 1.810 & 8.93 & 3.65 & 0.485 & 6.20 \\
$D4$   & 3.116 & 0.65 & 1.861 & 9.65 & 3.95 & 0.543 & 7.67 \\
\hline \\
\end{tabular}
\label{tableInitial}
\end{center}
\end{table}


\section{Initial stellar models}
\label{sec:model}

        As mentioned earlier, this paper is specially dedicated to the
study of the gravitational collapse of slowly and rapidly rotating
supramassive relativistic stars, in uniform rotation, that have become
unstable to axisymmetric perturbations.  Given equilibrium models of
gravitational mass $M$ and central energy density $e_c$ along a sequence
of fixed angular momentum or fixed rest mass, the Friedman, Ipser \&
Sorkin criterion \hbox{$\partial M/\partial e_c= 0$} \cite{Friedman88}
can be used to locate the exact onset of the {\it secular} instability to
axisymmetric collapse. The onset of the {\it dynamical} instability to
collapse is located near that of the secular instability but at somewhat
larger central energy densities. Unfortunately, no simple criterion
exists to determine this location, but the expectation mentioned above
has been confirmed by the simulations performed here and by those
discussed in~\cite{Shibata99e}. Note that, in the absence of viscosity or
strong magnetic fields, the star is not constrained to rotate uniformly
after the onset of the secular instability and could develop differential
rotation. In realistic neutron stars, however, very intense magnetic
fields are likely to counteract this.

        For simplicity, we have focused on initial models constructed
with a polytropic EOS $p=K\rho^{\Gamma}$, choosing $\Gamma=2$ and
polytropic constant $K_{\rm ID}=100$ to produce stellar models that are,
at least qualitatively, representative of what is expected from
observations of neutron stars. More specifically, we have selected the
models with the following procedure: first we have identified nine models
having polar-to-equatorial axes ratio in the interval 0.65--1.0 and lying
on the line defining the onset of the secular instability (the dotted
line in Fig.~\ref{fig:Initial}). The models used as initial data have
then been derived from the secularly unstable ones after increasing the
central energy density by $5\%$, while keeping the same axis ratio. These
models were indeed found to be dynamically unstable~\cite{Baiotti04} and
we have indicated them here as $D0, D1,\ldots, D3.5, D4$ following the
convention introduced in paper I. Note that model $D0$ effectively
corresponds to a TOV star.

	The main properties of these models are summarized in
Fig.~\ref{fig:Initial}, which shows the gravitational mass as a function
of the central energy density for equilibrium models constructed with the
chosen polytropic EOS. The solid, dashed and dotted lines correspond
respectively to: the sequence of non-rotating models, the sequence of
models rotating at the mass-shedding limit and the sequence of models
that are at the onset of the secular instability to axisymmetric
perturbations. The dynamically unstable initial models used in the
collapse simulations are shown as circles.

A more quantitative description of the models is presented in Table
\ref{tableInitial}, which summarises the main equilibrium properties of
the initial models. The circumferential equatorial radius is denoted as
$R_e$, while $\Omega$ is the angular velocity with respect to an inertial
observer at infinity, and $r_p/r_e$ is the ratio of the polar to
equatorial coordinate radii. Other quantities shown are the central
rest-mass density $\rho_c$, the ratio of the angular momentum $J$ to the
square of the gravitational mass $M$, and the ratio of rotational kinetic
energy to gravitational binding energy $T/|W|$.

\section{Dynamics of the Collapse}
\label{doc}

\begin{figure*}
\centering
\includegraphics[angle=0,width=6.cm]{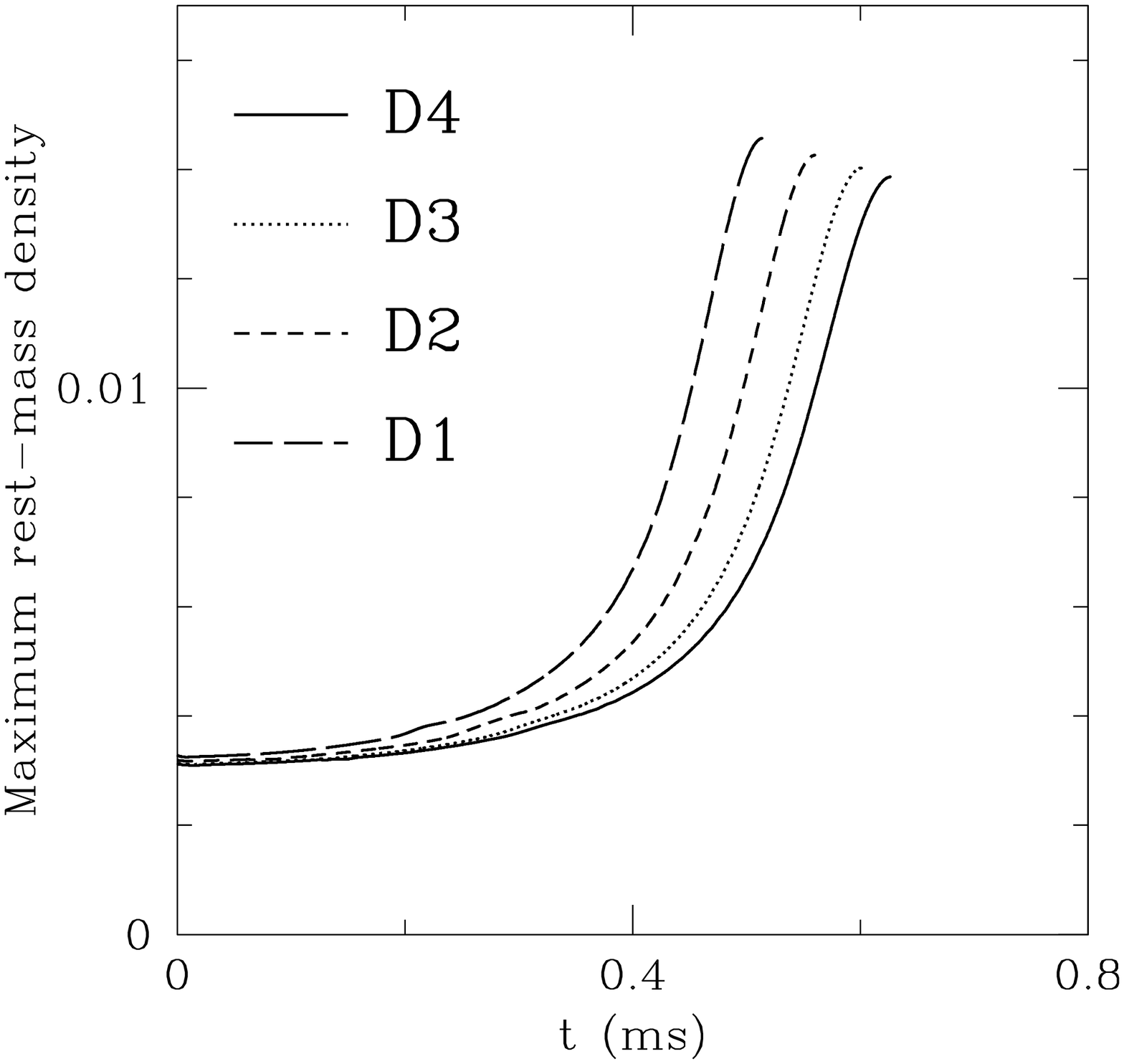} 
\hskip 0.25cm
\includegraphics[angle=0,width=6.5cm]{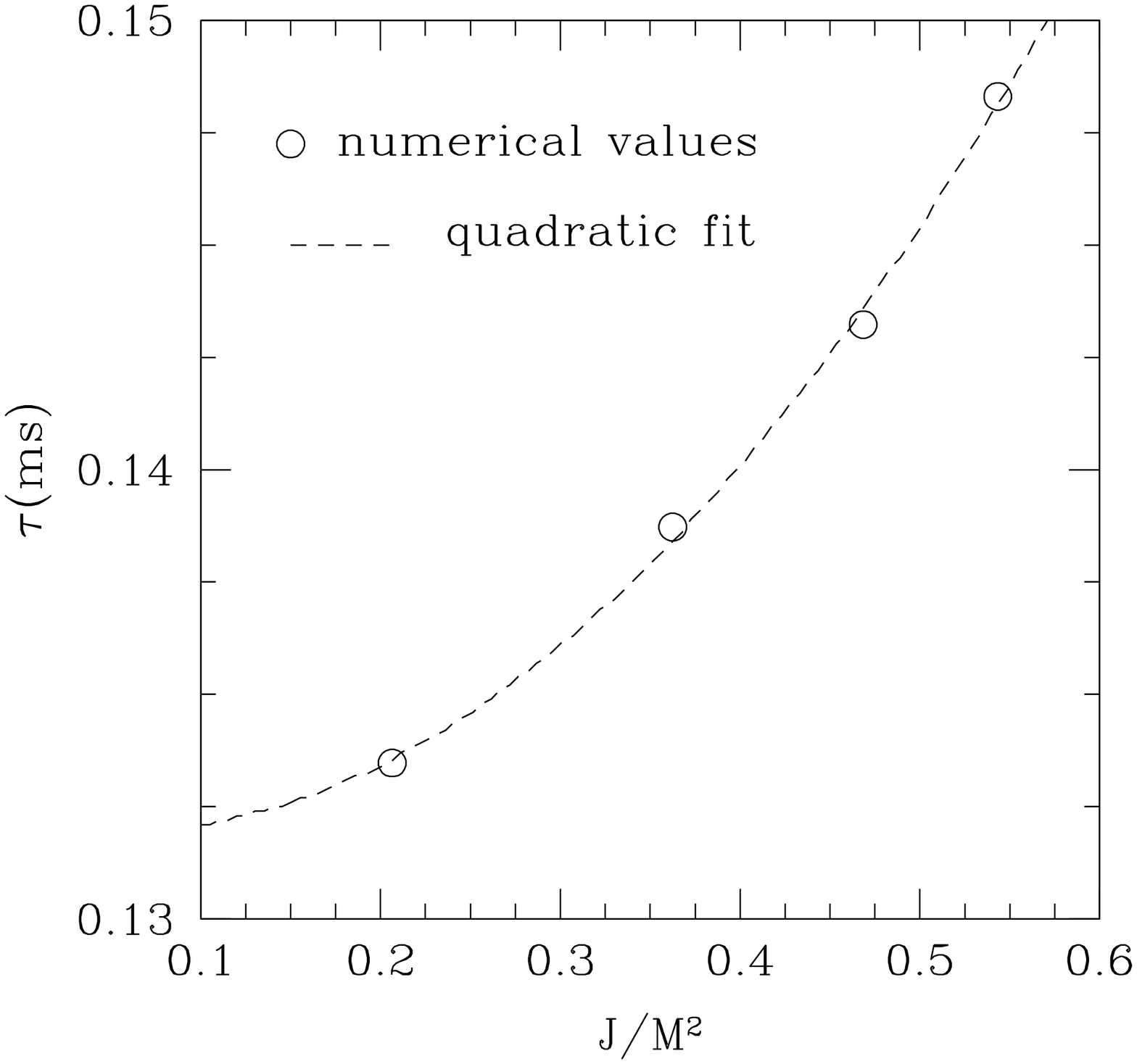} 
\caption{\textit{Left panel:} Time evolution of the maximum value of
  the rest-mass density for some representative models. \textit{Right panel:}
  exponential growth-time $\tau$ of the central density as a function
  of stellar rotation rate $J/M^2$. Indicated with open circles are
  the numerical values, while the dashed line is the very good fit
  obtained with a quadratic function in $J/M^2$.}
\label{fig:rhomax_evol} 
\end{figure*}

In paper I we have described in detail the dynamics of the matter and of
the apparent and event horizons during the gravitational collapse. Here,
we summarise previous results and provide additional details and
comparisons. 

We start by noting that although dynamically unstable models are expected
to collapse over a dynamical timescale, the collapse is traditionally
accelerated through the introduction of a small perturbation, either in
terms of an added radial velocity or through a slight and global
reduction of the pressure.  This is done, for instance, by using a
polytropic constant for the evolution $K$ that is 2\% smaller than the
one used to calculate the initial data $K_{\rm ID}$. We note that we do
not solve for the constraint equations once the initial perturbation is
introduced. This clearly produces a small error but, as shown in paper I,
after an initial transient lasting a couple of tenths of millisecond, the
constraint violation differs only of a few percent from the one measured
in a simulation in which the constraints had been re-solved.
A more detailed discussion of the influence of the type and amplitude of
the perturbations introduced on the waveforms emitted will be presented
in Section~\ref{votn}.

\begin{figure}
\centering 
\includegraphics[angle=0,width=8.5cm]{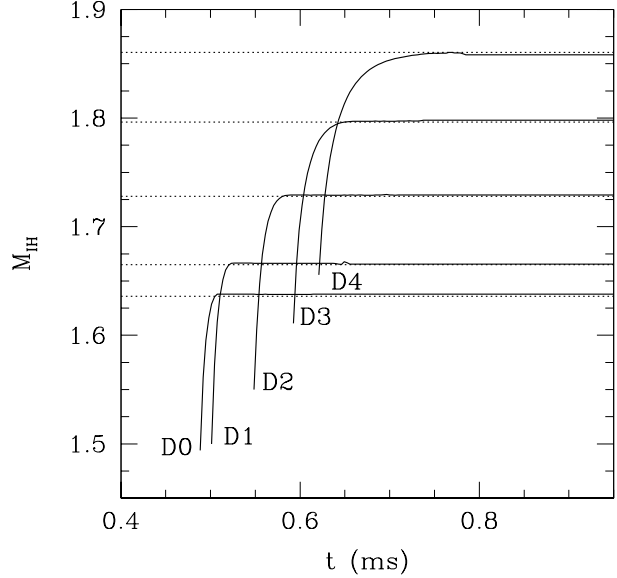}
\caption{Time evolution of the mass of the horizon in the dynamical
  horizon framework, for the different models (for clarity's sake, not
  all models are shown). The dotted lines represent the respective values
  of the ADM masses at the initial time, as computed after a
  compactification to infinity.}
\label{fig:IHs}
\end{figure}

Overall, all of the models, with different initial $J/M^2$ values, show 
similar dynamics as far as the bulk of the matter and the horizons are
concerned. The main differences across different models concern, instead,
the dynamics of the matter around the equatorial plane and the
surface of the star. Of course, models with higher $J/M^2$ are initially
more flattened and their oblateness increases as the collapse proceeds,
leading during the collapse to the temporary formation of a disc-like
configuration, which is however unstable and is rapidly accreted
({\it cf.} Figs.~5 and 6 of paper I).

As a good representative quantity, we show in the left panel of
Fig.~\ref{fig:rhomax_evol} the time evolution of the maximum value of the
rest-mass density for some of the models. Clearly all curves show an
exponential growth of the type $\rho = \rho_0 + A {\rm
exp}[(t-t_0)/\tau]$, where $\rho_0$ and $t_0$ refer to the initial
values, suggesting that the growth-time increases with the rate of
rotation, as a result of the increased centrifugal support. This is made
more clear in the right panel of Fig.~\ref{fig:rhomax_evol}, which shows
the exponential growth-time $\tau$ as a function of the rotation
rate. The open circles represent the numerical values, while the dashed
line is the excellent fit obtained with a quadratic function of $J/M^2$
with coefficients $c_0=0.13208,\ c_1=-0.00655,\ c_2=0.06642$, where $c_i$
is the coefficient of the term of order $(J/M^2)^i$. The increase of the
growth-time with the rotation rate is simple to explain in terms of the
increased centrifugal support that rapidly rotating models have and its
quantification represents an important result, being the first estimate
of the growth-time for the dynamical instability to axisymmetric
perturbations as computed in full General Relativity and for rapidly
rotating stars.

The evolution of another representative quantity is presented in
Fig.~\ref{fig:IHs}, which shows the behaviour of the masses of the
isolated horizons compared to the initial values of
the ADM masses computed at spatial infinity. As in paper I, the figure
shows that the mass of the newly formed black hole is measured very
accurately, and with an error, at the resolutions we have used here, of a
couple of percent only, when compared with the expected value of the
ADM mass.  Furthermore, this error is indeed comparable with the error
coming from the use of a finite-size domain and is of about one
percent (see paper I for more details).


\section{Extracting in the wave zone}
\label{eitwz}

The simulations presented in paper I, as well as other works ({\it
  e.g.}\ refs.~\cite{Shibata99e,Shibata:2003iy,Duez04hydro}), made use
of numerical grids with uniform spacing.  This, together with the
presently available computational resources, has initially forced us
to place the outer boundary of our computational domain in the {\it
  near zone}, {\it i.e.} in regions of the spacetime where the
gravitational waves have not yet reached their asymptotic form, which
instead happens in what is usually referred to as the {\it wave
  zone}. Under these constraints, the data on the gravitational
waveforms that we extract through gauge-invariant perturbative
techniques (see~\cite{Rezzolla99a,Nagar05,Baiotti06c} for details)
does not provide interesting information besides the obvious change in
the quadrupole moment of the background spacetime.

\begin{figure*}
\centering
\includegraphics[angle=0,width=6.4cm]{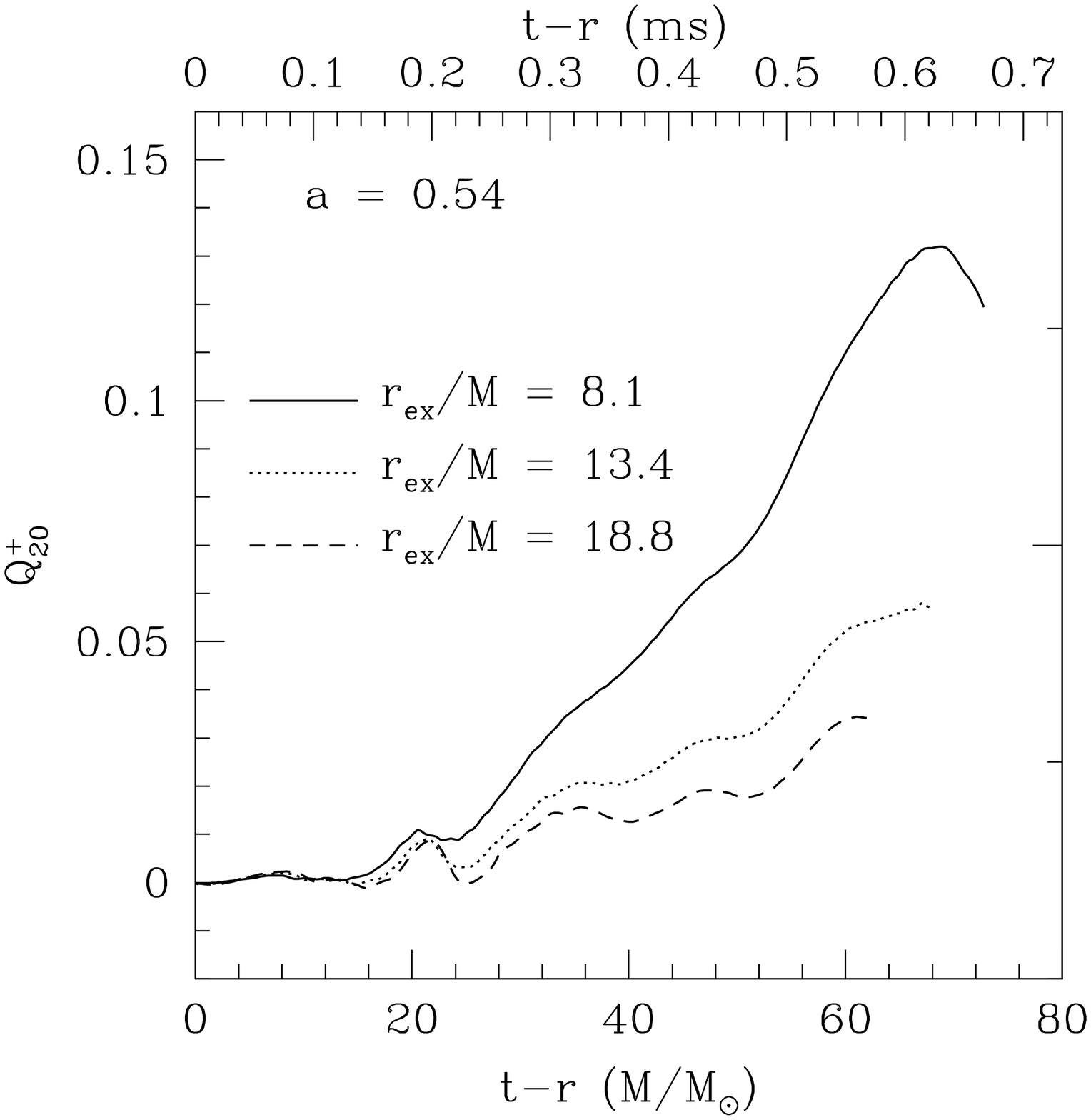} 
\includegraphics[angle=0,width=6.4cm]{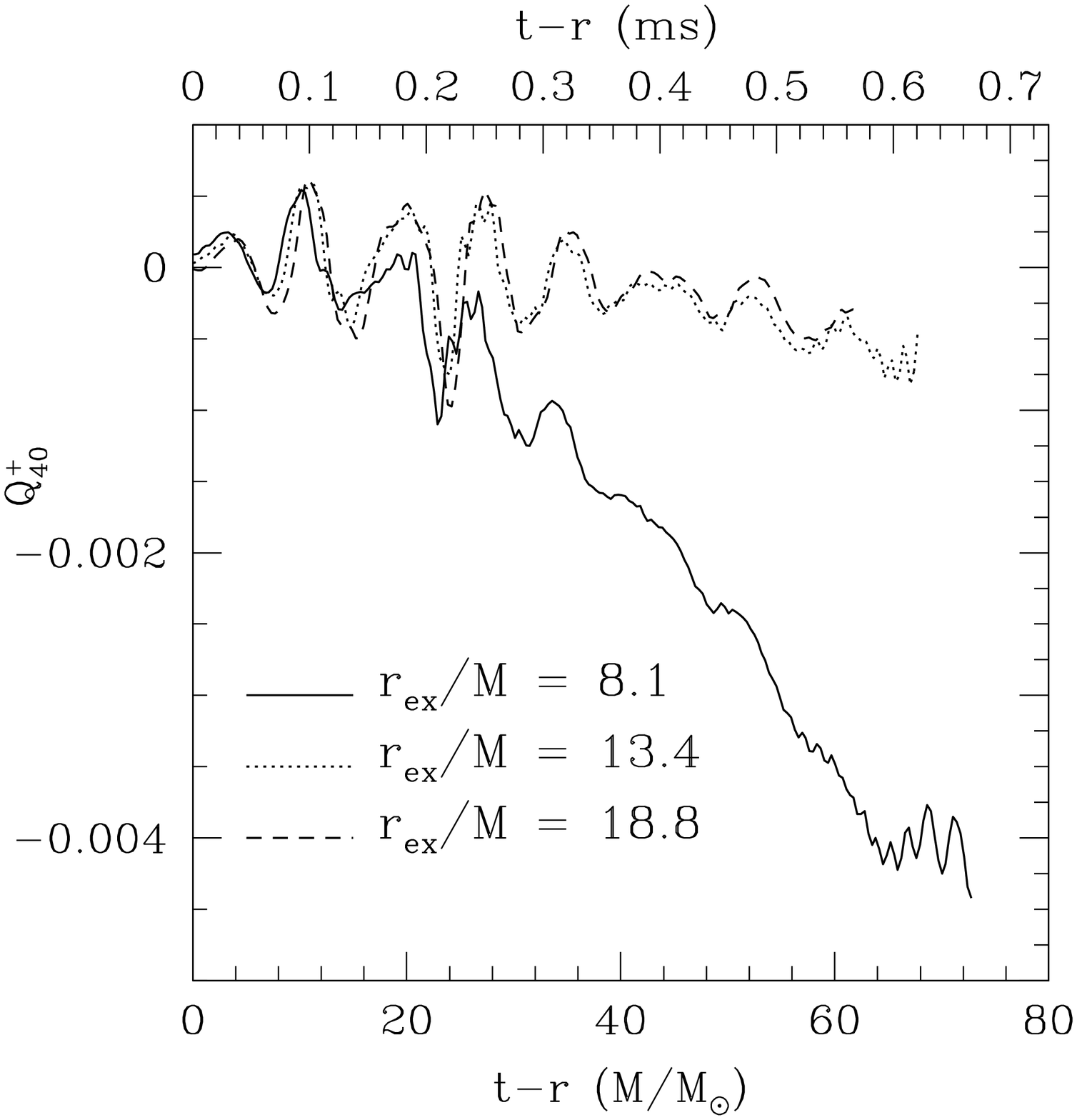}  
\caption{Gravitational wave extraction at short distances: waveforms of
  the even-parity metric perturbations $Q^+_{20}$ (left panel) and
  $Q^+_{40}$ (right panel) as functions of retarded time (shown both in
  ms and solar-mass units) for model $D4$ evolved on a uniform
  grid. Different lines refer to different extraction distances,
  expressed in M in the legend and corresponding, respectively, to
  coordinate radii 1.6, 2.6 and 3.6 times the initial coordinate stellar
  equatorial radius $R_{\,*}$.}
\label{fig:20_40_at_15} 
\end{figure*}

This is illustrated in Fig.~\ref{fig:20_40_at_15}, where we show the
the even-parity metric perturbations $Q^+_{20}$ (left panel) and
$Q^+_{40}$ (right panel) as functions of retarded time (shown both in
ms and solar-mass units) extracted at coordinate radii 1.6, 2.6
and 3.6 times the initial coordinate stellar equatorial radius,
$R_{\,*}$, 
or
equivalently at distances $8.1M$, $13.4M$, $18.8M$. These extraction
2-spheres are clearly not far enough out to be in the wave
zone. Indeed, we see (left panel) that the waveforms for the $\ell=2$
mode compared at the same retarded time do not overlap, as they should
if they were computed in the wave zone, since the invariance under a
retarded-time scaling is a property of the solutions of a wave
equation. The overlapping for the $\ell=4$ mode waves (right panel) is
slightly better and although quite noisy they show a wave-like
behaviour. Clearly this is possible because the higher-frequency waves
have a shorter wavelength and, less influenced by the secular changes
of the metric, reach the wave zone at smaller radii. However, also in
this case there clearly are secular variations of the waveforms that
are probably related to the dynamics of the gravitational field in the
near zone. We also note that the amplitude of the $\ell=4$ mode is
much smaller (one or two orders of magnitude) than that of the
$\ell=2$ mode, so one has to look primarily at the latter mode to
ascertain whether wave extraction has been performed successfully.

Using the mesh-refinement setup discussed in
Sect.~\ref{sec:mesh_refinement}, we were able to place the numerical
boundary of our coarsest grid much farther out. For our fiducial
simulation we use an outer boundary located at $\sim 160 M$ from the
central object. Because of the approximate boundary
conditions employed, at some time in the simulation numerical errors
reflected from the outer boundary arrive back in the central
high-gradient zone where they excite numerical instabilities which
are not cured by the small amount of dissipation applied. Given the
relatively short duration of the collapse, however, this does not
represent a serious problem and it is always possible to place the
outer boundary far enough out so that its influence is delayed to a time
when the largest part of the gravitational-wave emission has already
taken place. This boundary distance is indeed around $160 M$ for all
of the models studied here. Outer-boundary locations placed farther out
have not produced significant differences in the waveforms nor on the
emitted energy. In particular, comparing simulations of model $D4$ with
outer boundary located at $160M$ and at $320M$, the maximal pointwise
relative difference between correspondent values of the $\ell=2$ mode
is below 1\% (but the average difference is about 0.1\%) and the
relative difference in the emitted energy is below 0.5\%.

Finally, we note that our extraction 2-spheres are not located near
the outer boundary but, rather, around $50M$ from the origin and thus
at a distance which is about four times larger than the gravitational
wavelength. This distance is a good compromise between being far
enough in the wave zone and far enough from the outer boundary, from
where numerical contamination may come. We note that a similar
choice ({\it i.e.} extraction at $50M$) was made in~\cite{Stark85}.

\section{Variations on the theme: factors influencing the waveforms}
\label{votn}

A fundamental prospect of the world-wide effort dedicated to the
construction and planning of gravitational-wave detectors is that of
opening a ``new window'' on the universe through which we may observe details
of compact objects which would not be accessible through other
astronomical observations. As a step towards gravitational-wave
astronomy and the characterization of the sources through the features
of their gravitational radiation, we now discuss how the waveforms
computed here can provide important information on the physical
properties of the collapsing star. More specifically, we do this by
considering how the form and amplitude of the waves are influenced by
factors such as the rotation rate of the collapsing star or the type
and amplitude of the initial perturbations.

\begin{figure*}
\centering
\includegraphics[angle=0,width=5.9cm]{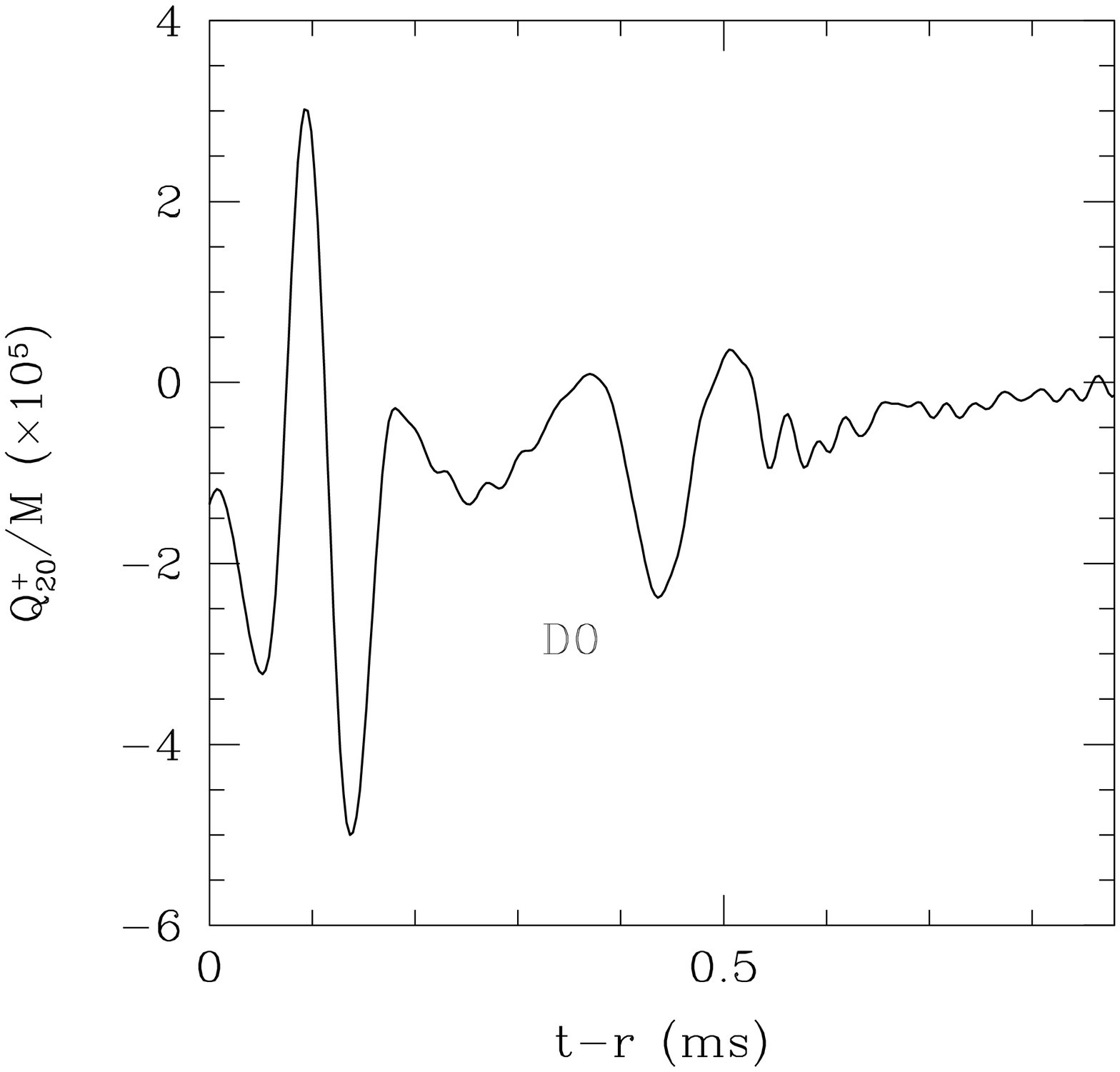} 
\hskip 1.0cm
\includegraphics[angle=0,width=5.9cm]{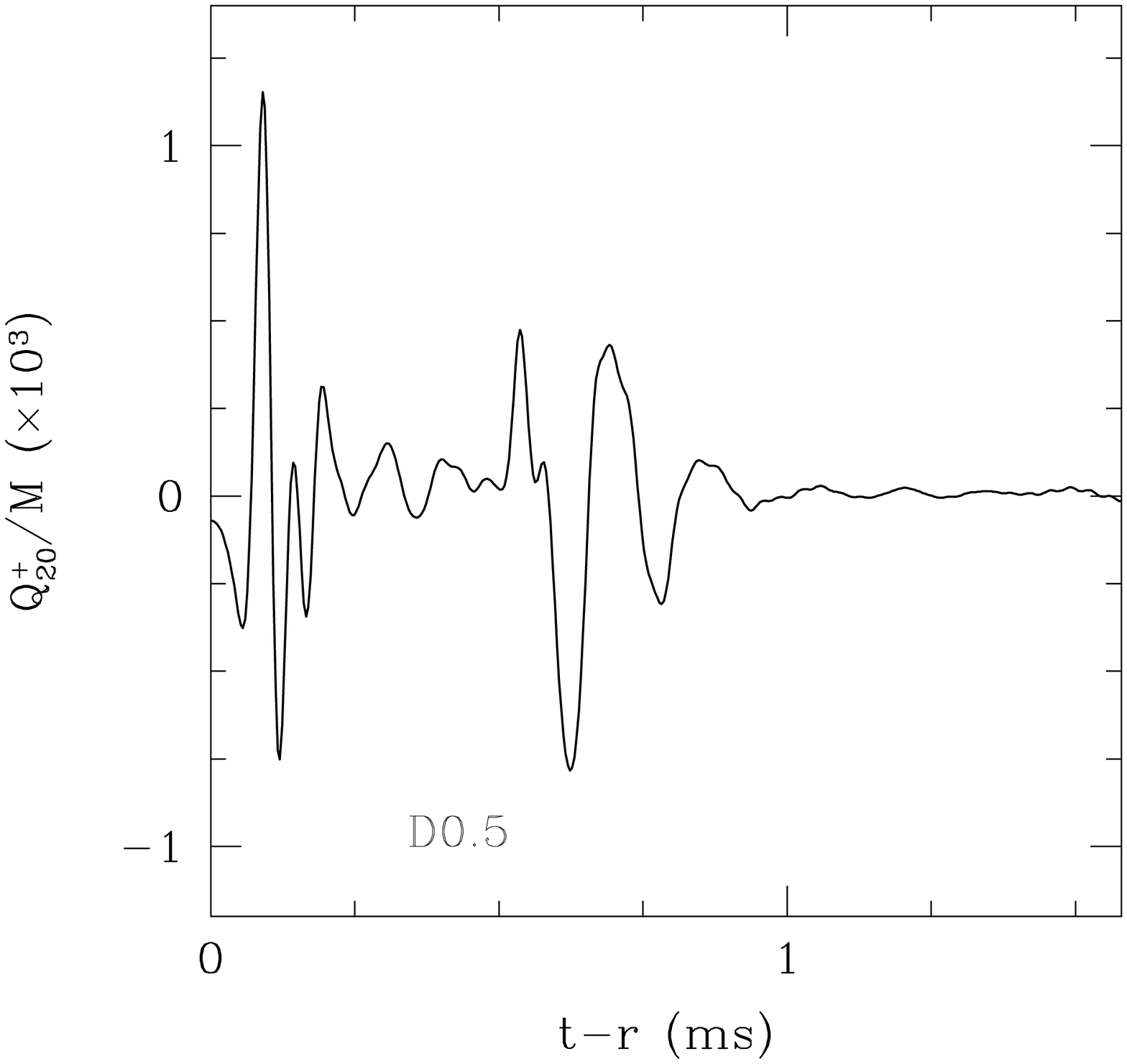}
\vskip 0.25cm
\includegraphics[angle=0,width=5.9cm]{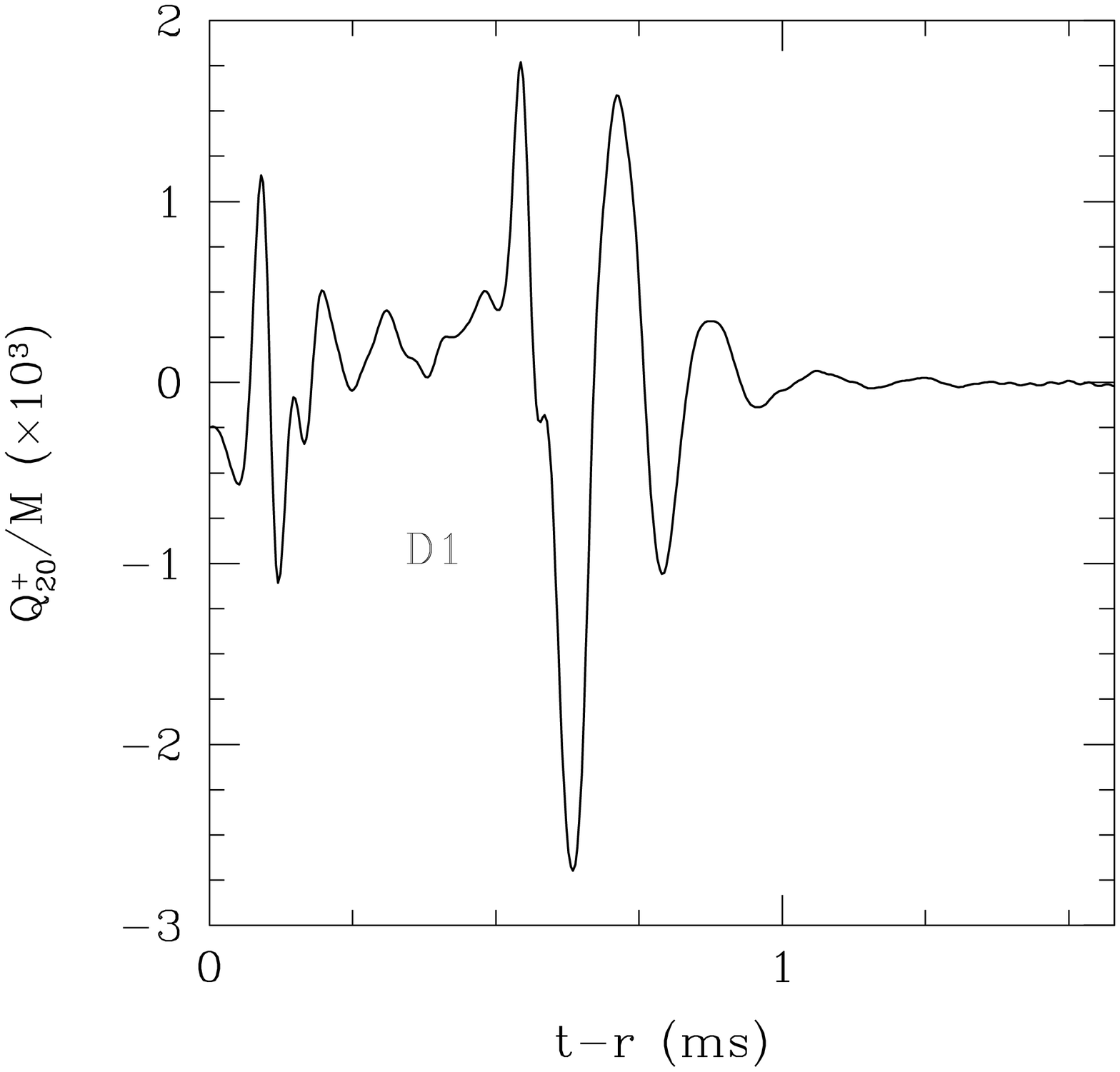} 
\hskip 1.0cm
\includegraphics[angle=0,width=5.9cm]{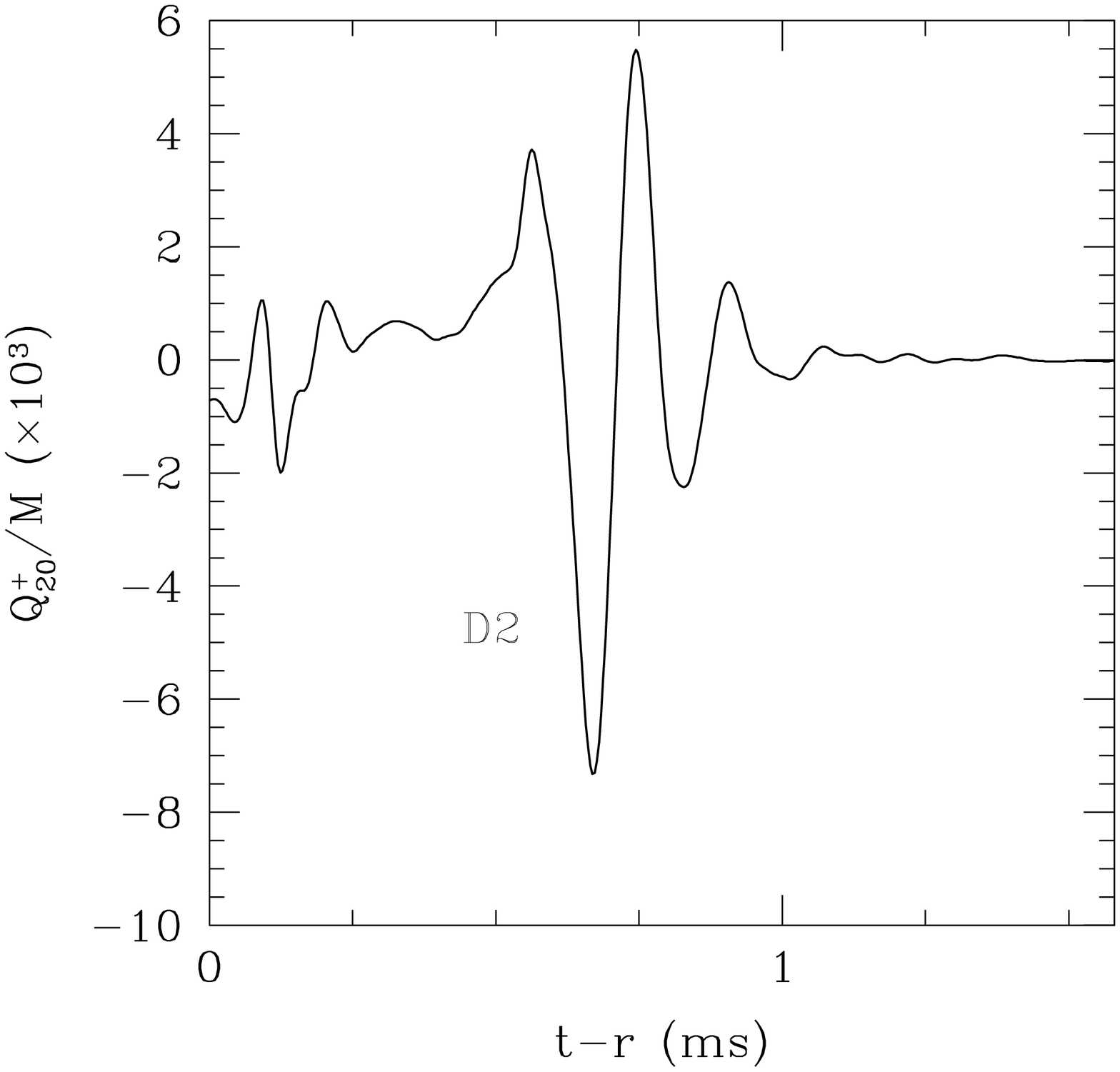} 
\vskip 0.25cm
\includegraphics[angle=0,width=5.9cm]{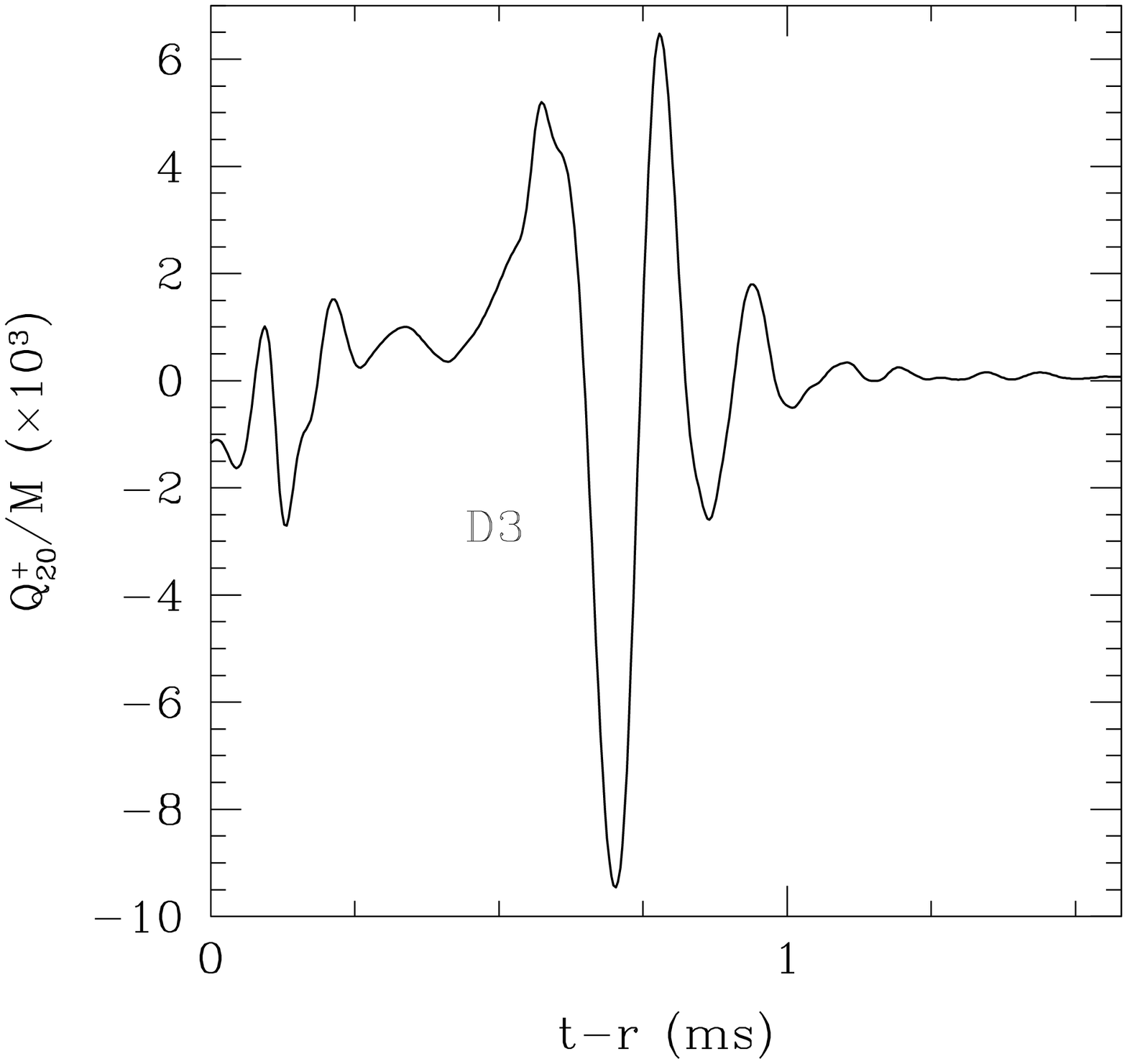} 
\hskip 1.0cm
\includegraphics[angle=0,width=5.9cm]{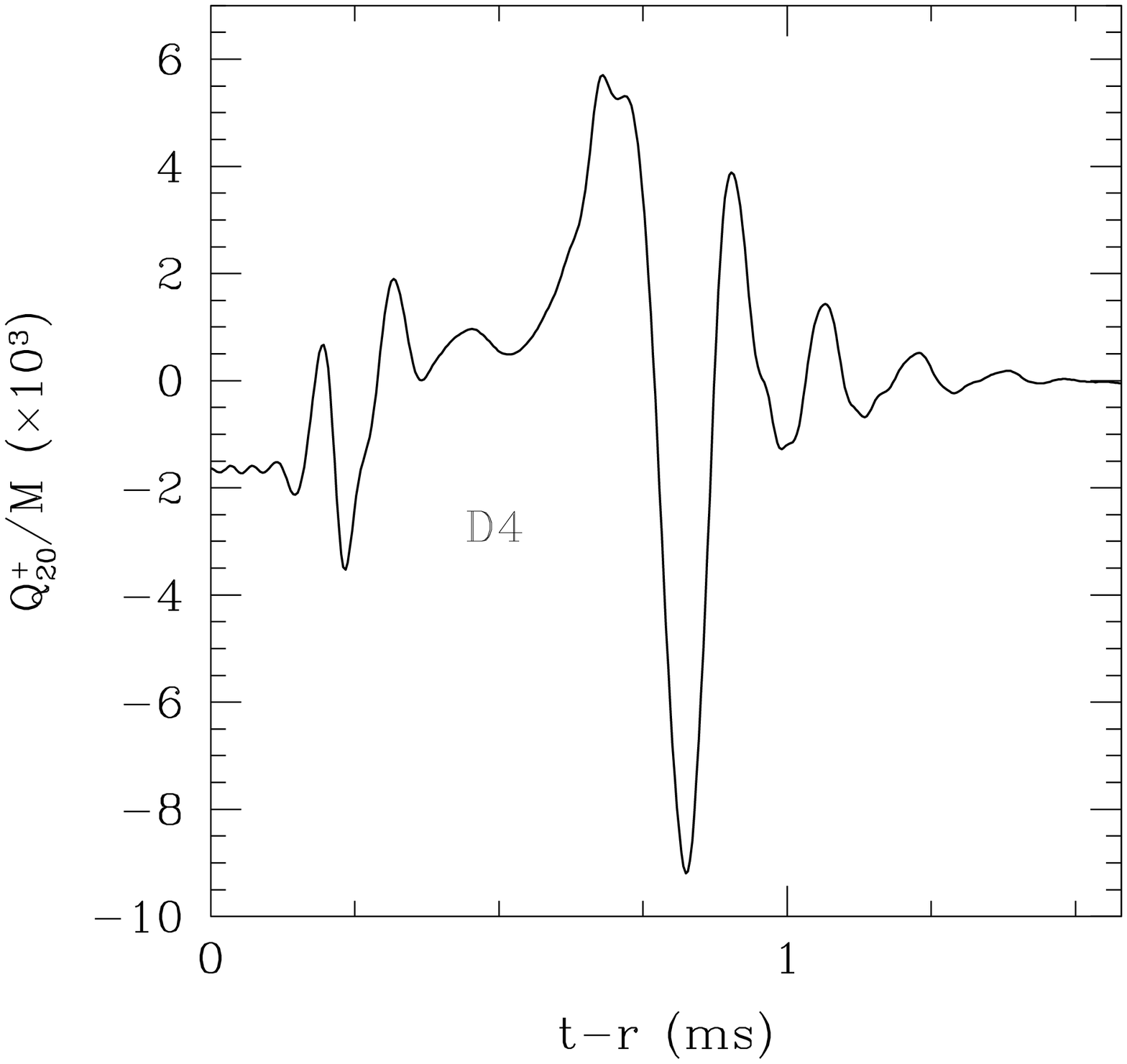}
\caption{$Q^+_{20}$ measured at $\sim 50~M$ in simulations of some of the
  models $D0-D4$, all with an initial pressure depletion of 2\%.}
\label{fig:Q20_vs_rotation}
\end{figure*}


\subsection{The Role of Rotation}
\label{ror}

Assessing the role that the stellar rotation rate has on the
emitted gravitational radiation is particularly simple in the case of
uniformly rotating stars, as all models can be selected so as to
differ only in the value of the angular velocity $\Omega$ after having
fixed either the central energy density or the gravitational
mass. Here, however, we consider the role of rotation along the
sequence of dynamically unstable models that we have discussed in the
previous Sections. More specifically, we show in
Fig.~\ref{fig:Q20_vs_rotation} the waveforms computed for some of the
simulated models when the initial model has been induced to collapse
through a reduction of about 2\% for the initial pressure support.

The waveform reported in the upper left panel is at least two orders
of magnitude smaller than any other waveform presented in
Fig.~\ref{fig:Q20_vs_rotation}, because it refers to the non-rotating
star $D0$ and should, at least in principle, be exactly zero. Model
$D0$, however, is not exactly a spherical star but rather a Cartesian
approximation of a spherical star at the level of resolution
considered here. Hence, the gravitational-wave signal in the upper
left panel should not be considered as an intrinsic error but, rather,
as a measure of the overall accuracy of our evolution code and
extraction technique.

\begin{figure*}
\centering
\includegraphics[angle=0,width=6.3cm]{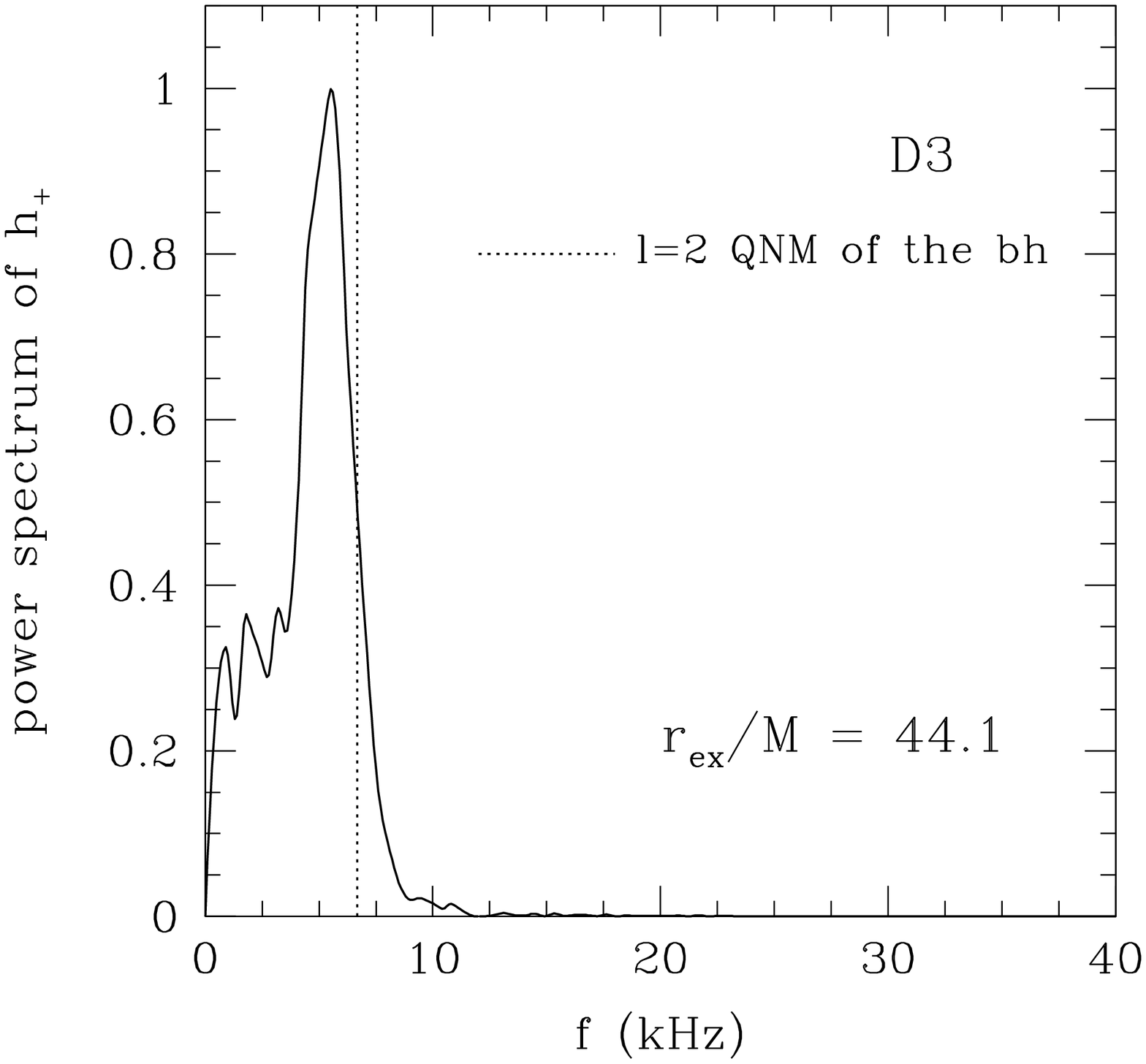}
\hskip 0.25cm
\includegraphics[angle=0,width=6.3cm]{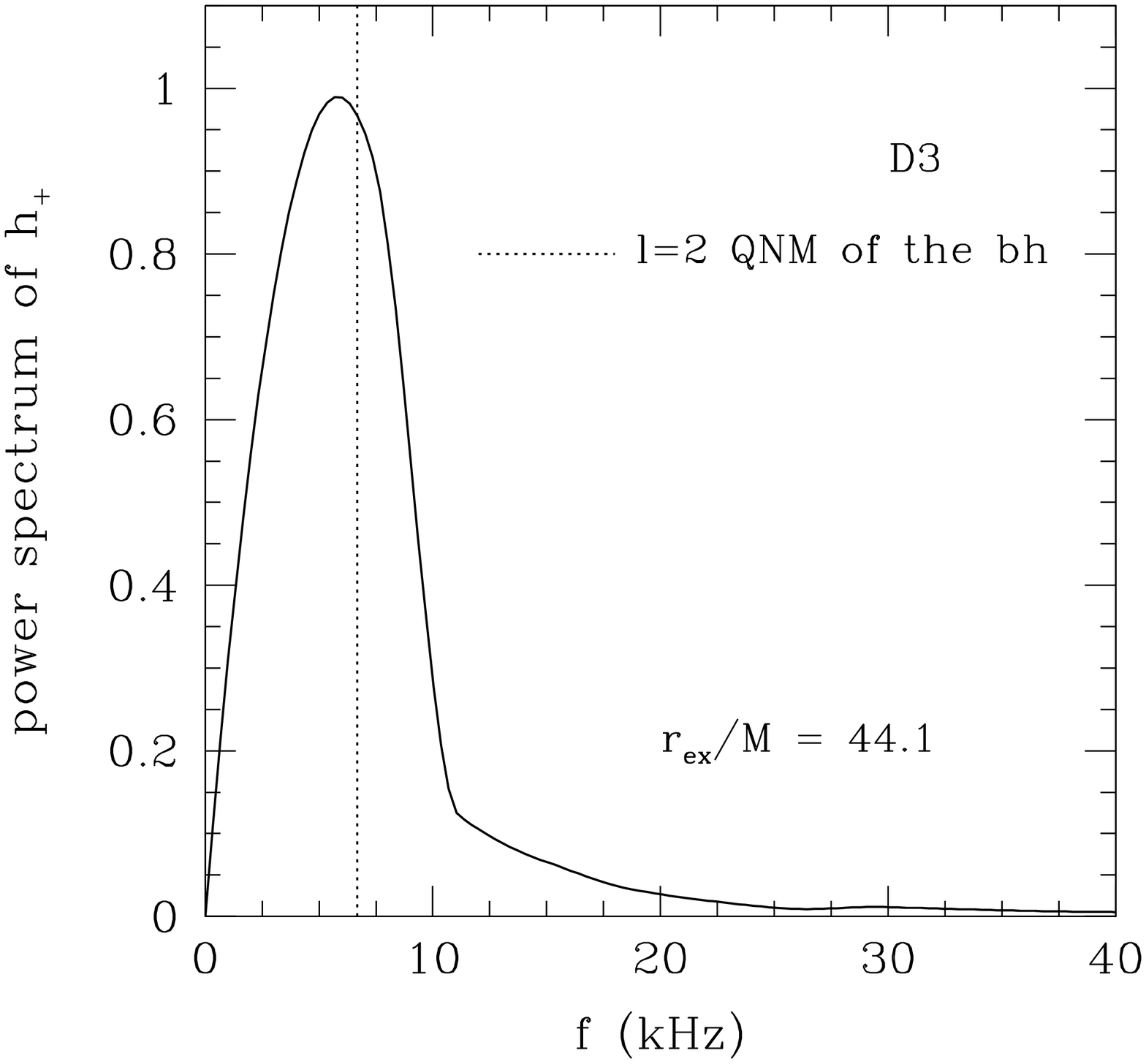}
\vskip 0.25cm
\includegraphics[angle=0,width=6.3cm]{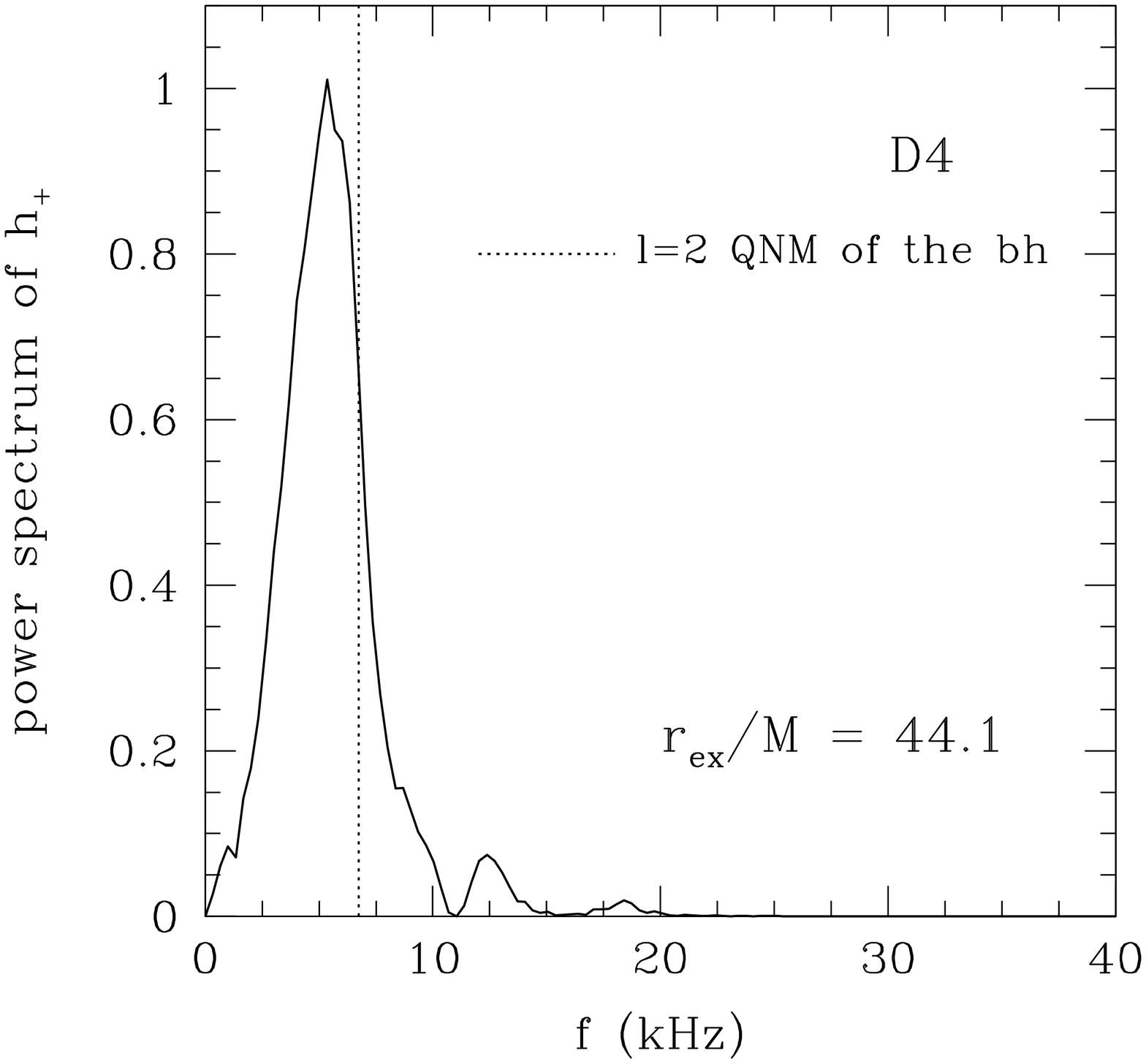}
\hskip 0.25cm
\includegraphics[angle=0,width=6.3cm]{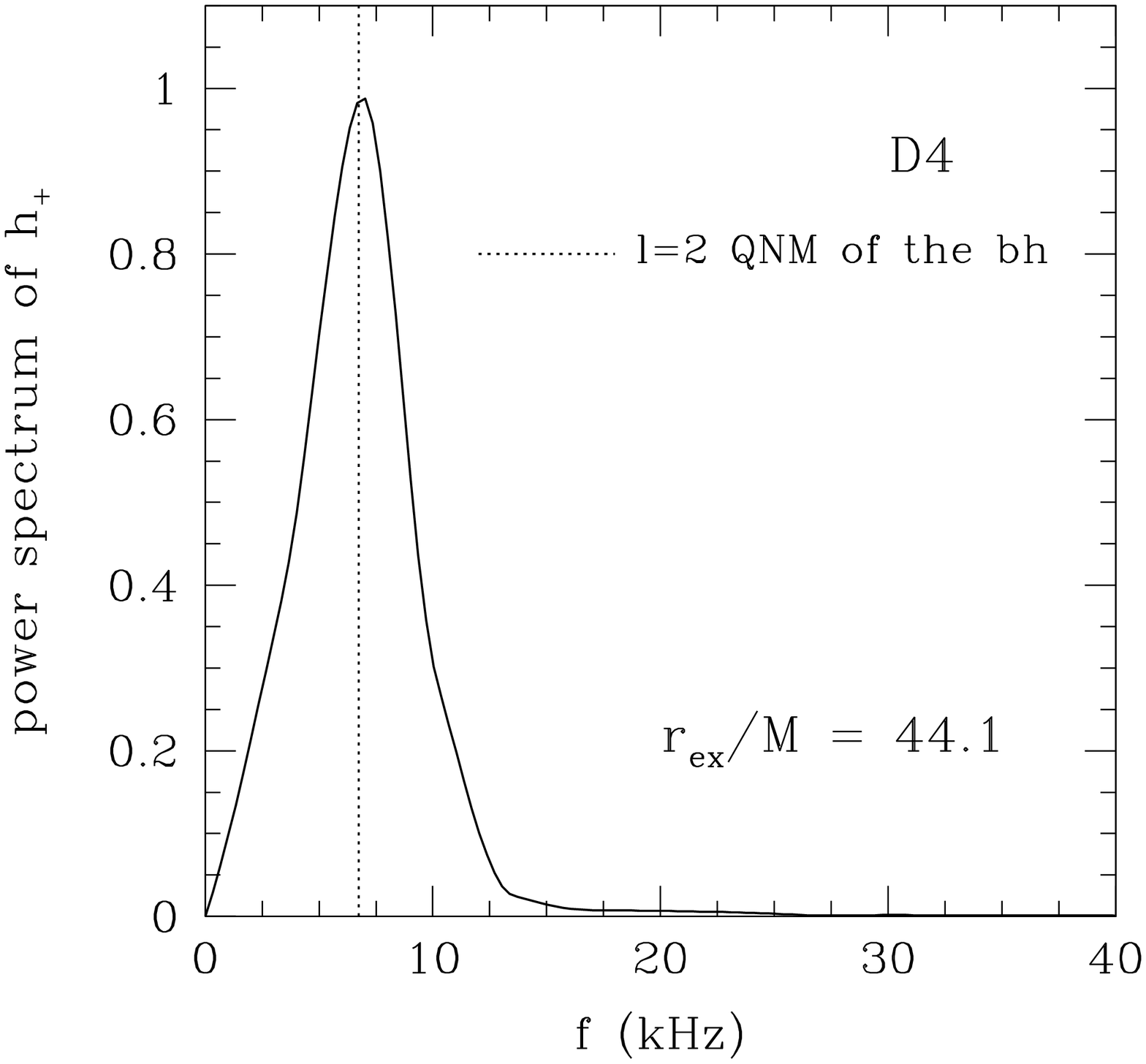}
\caption{Power spectra of $h_+$ measured at $\sim 50~M$. All panels
  refer to initial data without any added perturbation. Top left:
  complete extracted signal for $D3$; top right: only ring-down signal
  for $D3$; bottom left: complete signal for $D4$; bottom right: only
  ring-down signal for $D4$.}
\label{fig:PSD}
\end{figure*}

A rapid look at the waveforms in Fig.~\ref{fig:Q20_vs_rotation} is
sufficient to realize that the amount of initial rotational velocity
does influence both the amplitude and the form of the emitted
gravitational radiation. A more detailed discussion of this in terms
of the energy-efficiency and of the spectral properties of the signal
will be presented in Sect.~\ref{eead}. Here, however, it is sufficient
to underline that while the form of the signal does not vary
considerably, its amplitude changes by more than two orders of
magnitude over the range of possible rotations considered.

An interesting feature which is common to all the waveforms reported
in Fig.~\ref{fig:Q20_vs_rotation} is the presence of a high-frequency
signal between 0 and 0.25 ms and whose amplitude does not change
appreciably with rotation. As we will discuss in the following
Section, this initial and spurious burst of radiation is most likely
the signature of a perturbation in the star which is further amplified
by the reduction in pressure. Finally, we note that in all cases
considered the \textit{complete} signal has been collected, starting
from the initial spurious burst at the beginning of the collapse and up to
the ring-down phase of the black hole. After this, the extracted
signal becomes essentially constant until the numerical error produced
at the outer boundary reaches the region of the spacetime where the
fields are rapidly varying in space and destroys the solution (this is
not shown in the figure).

Finally, the different panels of Fig.~\ref{fig:PSD} offer information
about the properties of the waveforms that is complementary to the one
presented in Fig.~\ref{fig:Q20_vs_rotation}. More specifically, they
show the power spectral density (PSD) of the waveforms emitted by
models $D3$ and $D4$ in the absence of any initial perturbation. The
left panels, in particular, show the PSD of the complete signal and
thus including also the initial spurious burst (see Sect.~\ref{ropg}
for a discussion of this), while the right panels show the PSD of only
the final part of the waveforms, namely the one produced by the
ring-down of the black hole. Indicated with a vertical dotted line is
the corresponding frequency of the fundamental black hole quasi-normal
mode (QNM) as computed in~\cite{Berti06c}. As expected, the PSD of the
complete signal is rather narrow and shows a main peak around 6 kHz
and a series of smaller peaks at larger frequencies, related to the
initial burst of radiation and possibly a signature of the $w$ modes
of the perturbed star. The PSD of the QNM ringing, on the other hand,
is wider in frequency but very well matched with the expected
frequency of the fundamental QNM.

\begin{figure}
\centering
\includegraphics[angle=0,width=9.5cm]{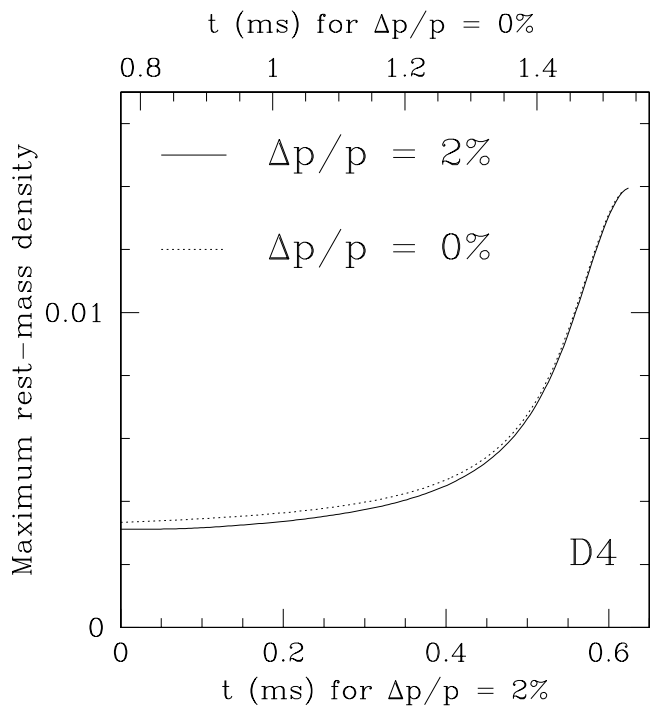}
\caption{Comparison of the time evolution of the maximum value of the
rest-mass density for model $D4$ with and without pressure depletion at the
initial time. The curve for the unperturbed case has been shifted in time
to offer a direct comparison. The time reference for the non-perturbed data
is reported on the upper axis.}
\label{fig:rhomax_nopert} 
\end{figure}


\subsection{The Role of the Pressure Perturbation}
\label{ropg}

As discussed in Sect.~\ref{sec:model}, it is customary in simulations
of collapse to rotating black holes to introduce a pressure
perturbation whose amplitude can be rather large (as
in~\cite{Stark85}, where the pressure support was decreased up to
99\%) or rather small (as in paper I, where the pressure support was
decreased by only 2\%). The rationale behind this approach is that the
introduction of the perturbation simply increases the amplitude of the
(only) unstable mode, hence triggering the instability and decreasing
the computational costs. As we will show below, this assumption is
correct only for very small perturbations and, quite on the contrary,
large-amplitude perturbations can have a strong impact on both the
dynamics of the matter (and hence of the horizons) and on the
gravitational waveforms.

We start by comparing in Fig.~\ref{fig:rhomax_nopert} the evolution of
the central rest-mass density for model $D4$ in the case in which a
pressure perturbation of 2\% is introduced (i.e. $\Delta p/p = 2\%$)
and when no explicit perturbation is introduced (i.e. $\Delta p/p =
0\%$). In this latter case, the fact that the model is already past
the secular instability limit and the presence of a small but nonzero
truncation error are sufficient to trigger the instability which leads
to a collapse (delayed with respect to the perturbed case). The two curves in
Fig.~\ref{fig:rhomax_nopert} are properly shifted in time so as to be
superposed and the upper $x$-axis is used to indicate the coordinate
time in the case of the unperturbed collapse.  It is apparent that in
this regime of linear perturbations the dynamics of the matter and
that of the horizons during the collapse (not shown here) are very
similar in both the perturbed and the unperturbed models.

\begin{figure*}
\centering
\includegraphics[angle=0,width=6.3cm]{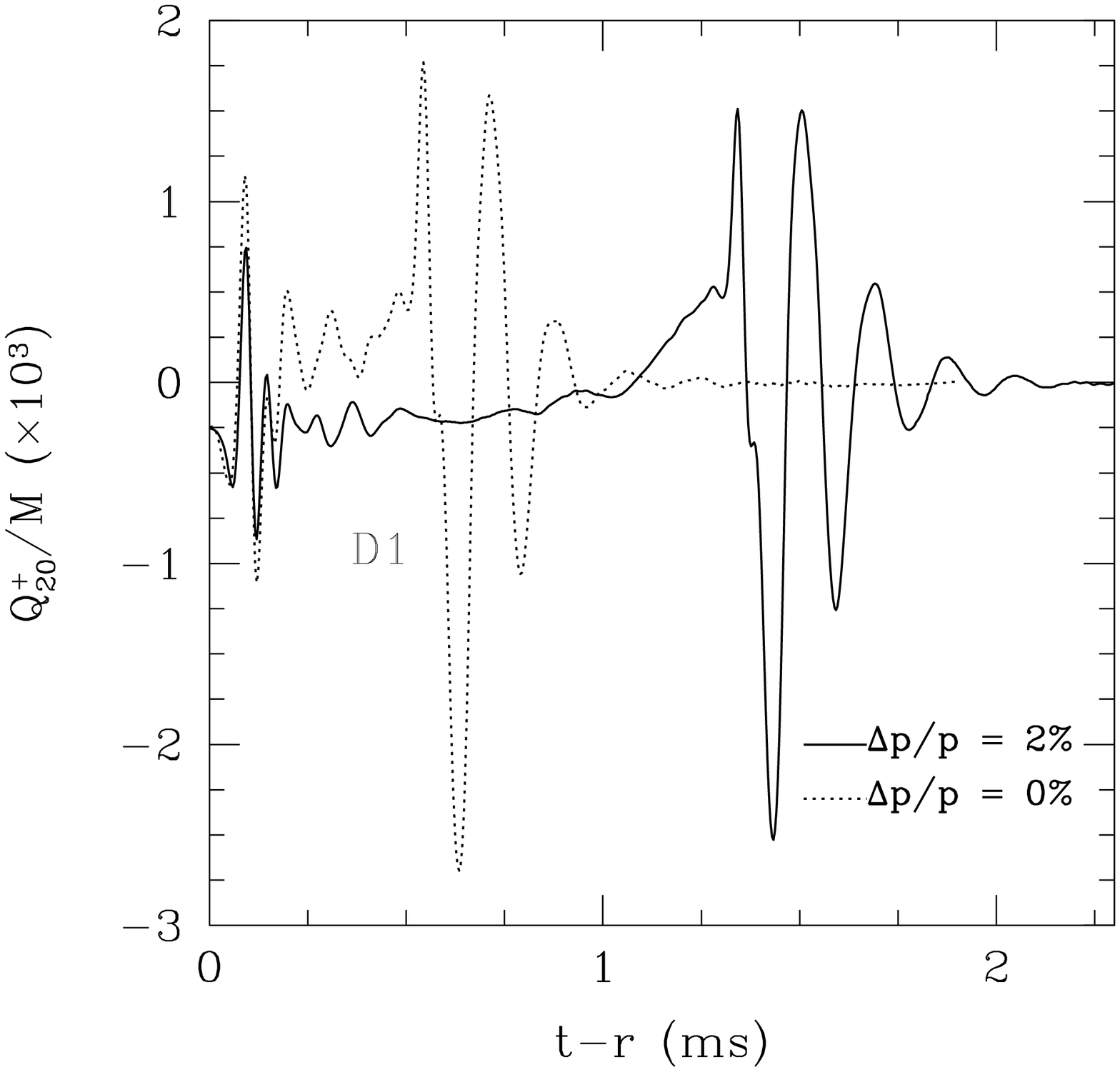}
\hskip 0.25cm
\includegraphics[angle=0,width=6.3cm]{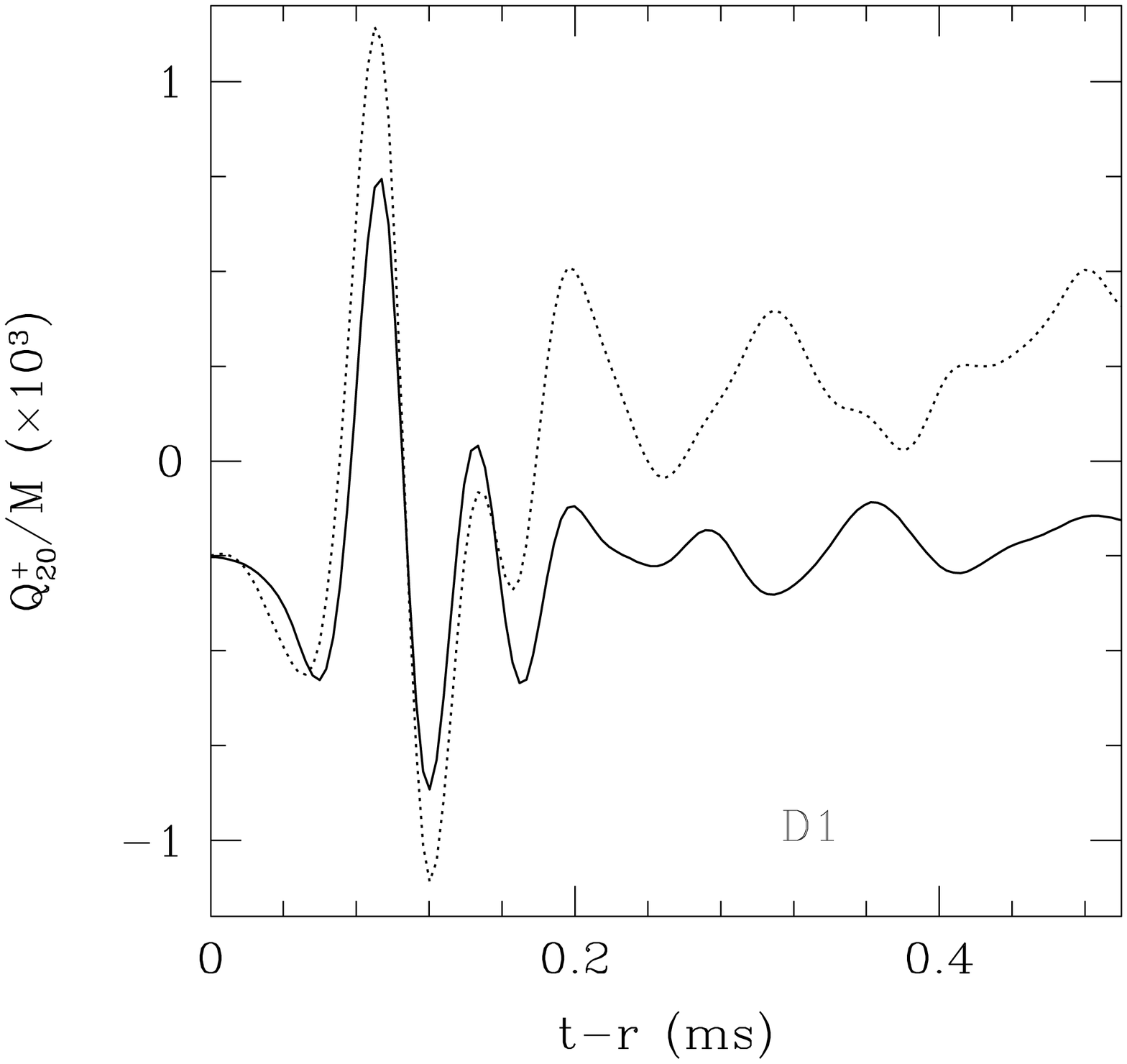}
\caption{\textit{Left panel:} Comparison of the $Q^+_{20}$ measured at
  $\sim 50~M$ in simulations of models $D1$ with an initial pressure
  depletion of 2\% (dotted line) and without any added perturbation
  (solid line). \textit{Right panel:} Magnification of the initial
  spurious burst.}
\label{fig:Q20_D1_Dp2+0}
\end{figure*}

A similar conclusion can be drawn when considering the
gravitational-wave emission and this is summarised in
Fig.~\ref{fig:Q20_D1_Dp2+0}, whose left panel shows the $Q^+_{20}$
waveforms for model $D1$ with the initial perturbation (dotted line;
{\it cf.}  Fig.~\ref{fig:Q20_vs_rotation}, center-left panel) and without
(solid line). Note that the two curves are not shifted in time and
thus the delay is effectively due to the smaller initial amplitude of
the unstable eigenmode when $\Delta p/p=0$. Note that not all of the
signal coming from the unperturbed model is shifted in time and indeed
also in this case an initial spurious burst of radiation is present
between and 0 and 0.25 ms, as highlighted in the right panel of
Fig.~\ref{fig:Q20_D1_Dp2+0}. As mentioned earlier, this signal
originates essentially from the truncation error introduced when
interpolating onto a Cartesian grid the initial stellar models which
are computed as equilibrium models in a code using spherical polar
coordinates~\cite{Stergioulas95}. As a result, it is always
present, with a form which is essentially independent of the stellar
rotation rate ({\it cf.} Fig.~\ref{fig:Q20_vs_rotation}), but with an
amplitude which can be further increased if the star is perturbed and
hence with a larger initial violation of the constraint equations.
This is very evident in the right panel of Fig.~\ref{fig:Q20_D1_Dp2+0}
which shows the two signals being well superposed in phase but also
having different amplitudes, with the one coming from the unperturbed
star being systematically smaller.

It is as yet uncertain whether this initial signal, albeit spurious,
reflects a consistent response of the star to a perturbation and can
therefore be associated to a $w$ mode~\cite{Kokkotas99a}. Preliminary
investigations in this direction seem to support the idea that the
gravitational signal between 0 and 0.25 ms does indeed correspond to a
$w$ mode (the signal does not converge away with resolution as shown
in Fig.~4 of~\cite{Baiotti06c}) and could therefore be used to extract
the eigenfrequencies of these modes in rapidly rotating stars which
are yet unaccessible to perturbative studies. However, further work is
needed to consolidate this conclusion.

\begin{figure}
\centering
\includegraphics[angle=0,width=6.2cm]{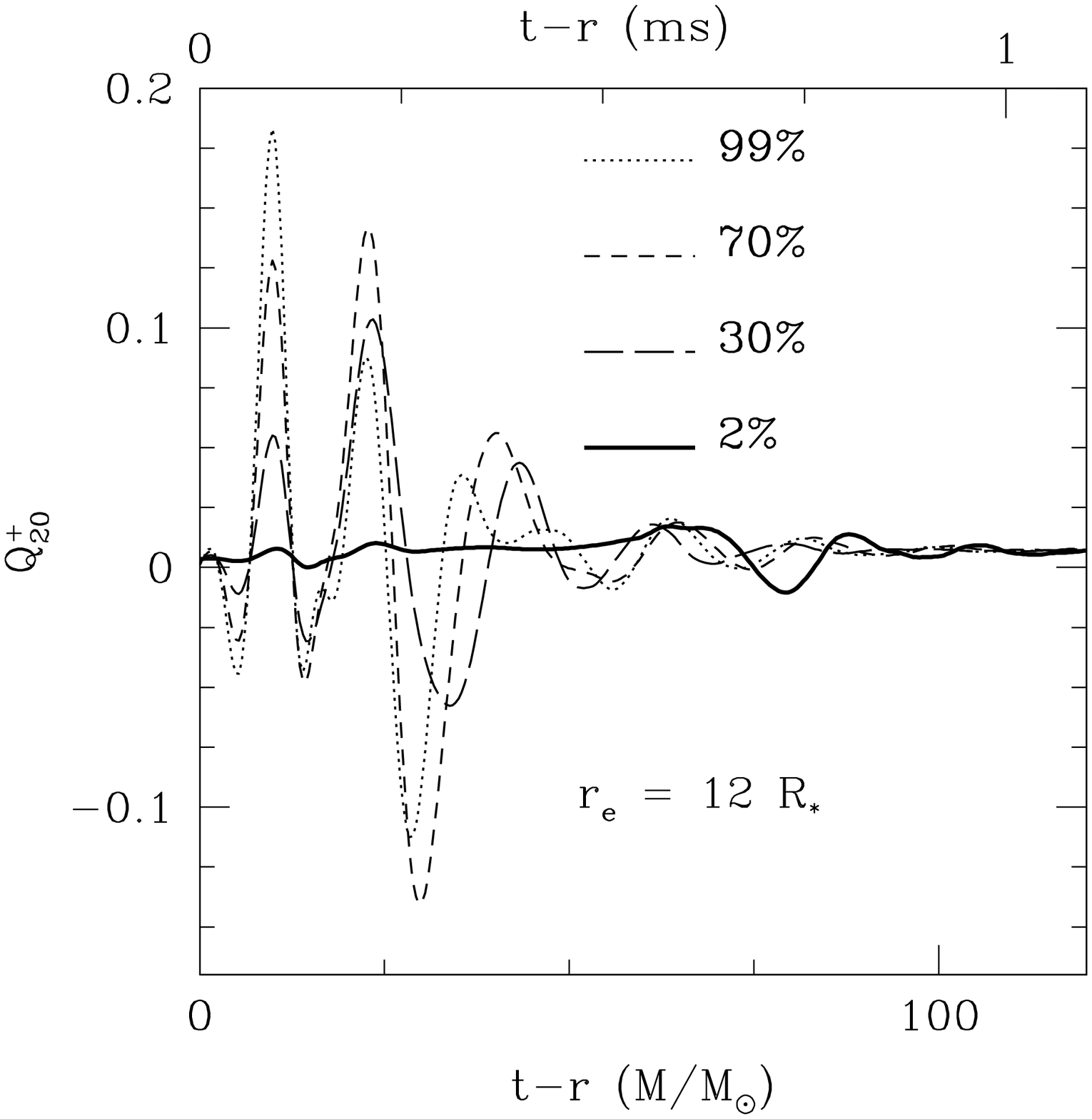} 
\hskip 0.5cm
\includegraphics[angle=0,width=6.2cm]{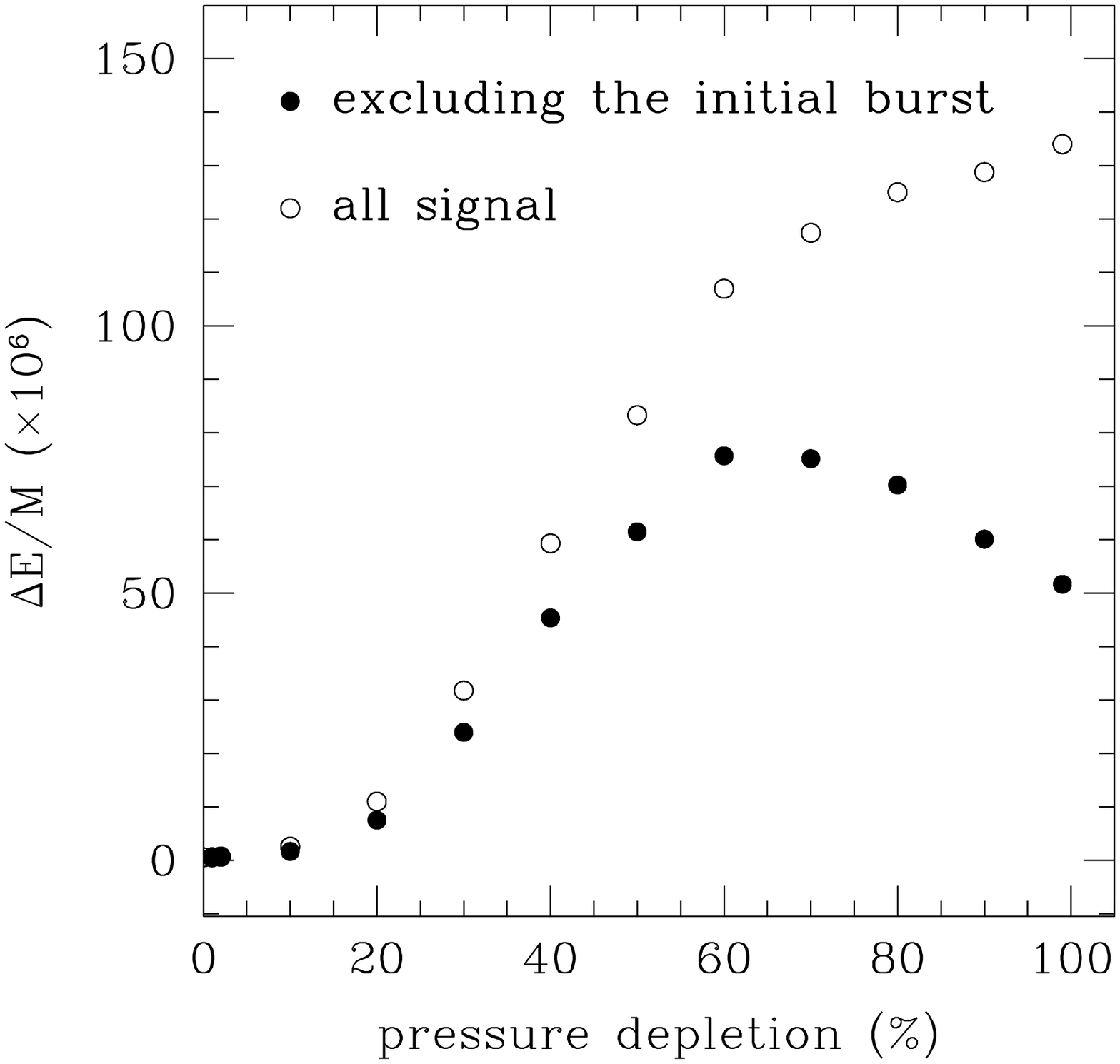} 
\caption{\textit{Left panel:} Comparison of $Q^+_{20}$ from
  simulations of model $D4$, differing in the amount of initial
  pressure depletion. \textit{Right panel:} Energy carried by the
  emitted gravitational waves during the collapse of model $D4$ with
  different percentages of initial pressure depletion. All
  measurements were performed at a coordinate distance of $\sim 50
  M.$}
\label{fig:Q20_vs_pressdep}
\end{figure}

We note that while introducing an initial \textit{small} perturbation
can serve to accelerate the matter-dynamics and that the latter is not
influenced noticeably, a \textit{large} pressure perturbation can
however lead to significantly different results and even to incorrect
interpretations on the efficiency of the gravitational-wave emission
during the collapse. This is quite apparent in
Fig.~\ref{fig:Q20_vs_pressdep}, whose left panel shows the changes in
the even-parity waveform $Q^+_{20}$ emitted during the collapse of
model $D4$ when this is subject to pressure depletions going from 2 to
99\%.  Clearly, as the pressure depletion is increased, the
gravitational collapse becomes much more rapid, asymptotically
becoming the one produced in the free-fall of an oblate distribution
of dust. The corresponding changes in the mass quadrupole from this
increasingly rapid collapse are also larger and thus the amplitude of
the gravitational-wave emission is also increased. It is not
surprising, therefore, that in these conditions it can easily reach
values comparable with the ones computed in~\cite{Stark85} who
were indeed using pressure reductions between 60 and 99\%.

In order to find a closer comparison with the values reported
in~\cite{Stark85} we show in the right panel of
Fig.~\ref{fig:Q20_vs_pressdep} the energy emitted in gravitational
waves as a function of the pressure depletion for model $D4$. Note,
however, that in the case of large pressure depletions, the collapse
is so rapid that it is very difficult, if possible at all, to
distinguish the initial-burst signal from the one produced by the
collapse. Indeed, although the height of the first peak of the signal
is closely related to the amplitude of the initial perturbation and
grows monotonically with it, this is not true for the other peaks, as
shown in the left panel of Fig.~\ref{fig:Q20_vs_pressdep}. As a
result, while the open circles are refer to the total signal, the
filled ones instead, show the emitted energy when the first peak in
the signal is not taken into account and hence without the initial
burst. It is apparent that as the pressure removal is increased, the
energy radiated increases, becoming about two orders of magnitude
larger than the one obtained in the absence of perturbations. Such
large values are in good agreement with those presented
in~\cite{Stark85} and induce us to conclude that the estimates made
there, although served as useful upper limits, were dominated by the
unrealistic dynamics of the matter.

Interestingly, the total energy emitted in gravitational radiation 
(when not including the initial burst) does not have a monotonic behaviour
with $\Delta p/p$ and two different factors may combine to yield this
effect. The first one is that as the pressure support is drastically
reduced the centrifugal support, that in model $D4$ plays an important
dynamical role, ceases to be relevant and the dust-like matter
collapses with only a small increase in the oblateness and hence in
the mass quadrupole. The second factor is that in the more rapid
collapse triggered by larger pressure depletions, the apparent horizon
is produced much earlier and hence a larger amount of radiation
remains trapped and cannot reach the observer. While of little
practical interest because of the extreme conditions of matter
involved, verifying these conjectures may provide important
information on the behaviour of the Einstein equations in a nonlinear
regime and deserves further investigations.

As a final remark, we note that an initial small perturbation also has
an obvious drawback when it comes to analysing the gravitational-wave
signal. The signal from the early burst of radiation, in fact,
combines with the stronger collapse signal and can lead to incorrect
estimates about the efficiency of the emission of gravitational
radiation during the collapse for slowly rotating models. In the case
of model $D1$, for instance, the energy contained in the initial burst
amounts to $\sim 30\%$ of the energy produced instead during the
actual gravitational collapse. For model $D4$, on the other hand, this
amounts only to $\sim 2\%$.


\subsection{Perturbations in the velocity}
\label{roip}

A pressure reduction is not the only possible perturbation that can be
introduced in order to induce the collapse of a star past the secular
stability limit. Another possibility, also used in the past
in~\cite{Bonazzola-etal-1996:spectral-methods-in-gr}, consists in
adding an inward-directed radial velocity to the equilibrium
configuration, which we have here done in terms of a radial velocity
of constant modulus $0.02$ throughout the star.

This different type of initial perturbation gives rise to a slightly
larger violation of the constraint equations at the initial time and
produces a collapse over a timescale which is comparable to that
resulting from a 10\% depletion of the pressure support. The
efficiency in the energy emission, on the other hand, can be much
larger as will be discussed in more detail in the following Section.


\section{Energy-Efficiency and Detectability}
\label{eead}

Determining the energy-efficiency in the emission of gravitational
radiation in fully nonlinear regimes of the Einstein equations is
particularly difficult as perturbative or post-Newtonian approaches
cannot be used reliably. The role that numerical-relativity
calculations can therefore play in this context is therefore
particularly valuable and it represents one of the goals of most
simulations. In addition, determining this efficiency in the case of
the gravitational collapse to a black hole is made more difficult by
the intrinsic weakness of the system which looses only a small
fraction of its binding energy to gravitational radiation. In the case
of binary black hole calculations, in contrast, the efficiency can
easily reach a few percent even in the simplest scenario of
non-spinning, equal-mass binaries.

\begin{figure}
\centering \includegraphics[angle=0,width=6.2cm]{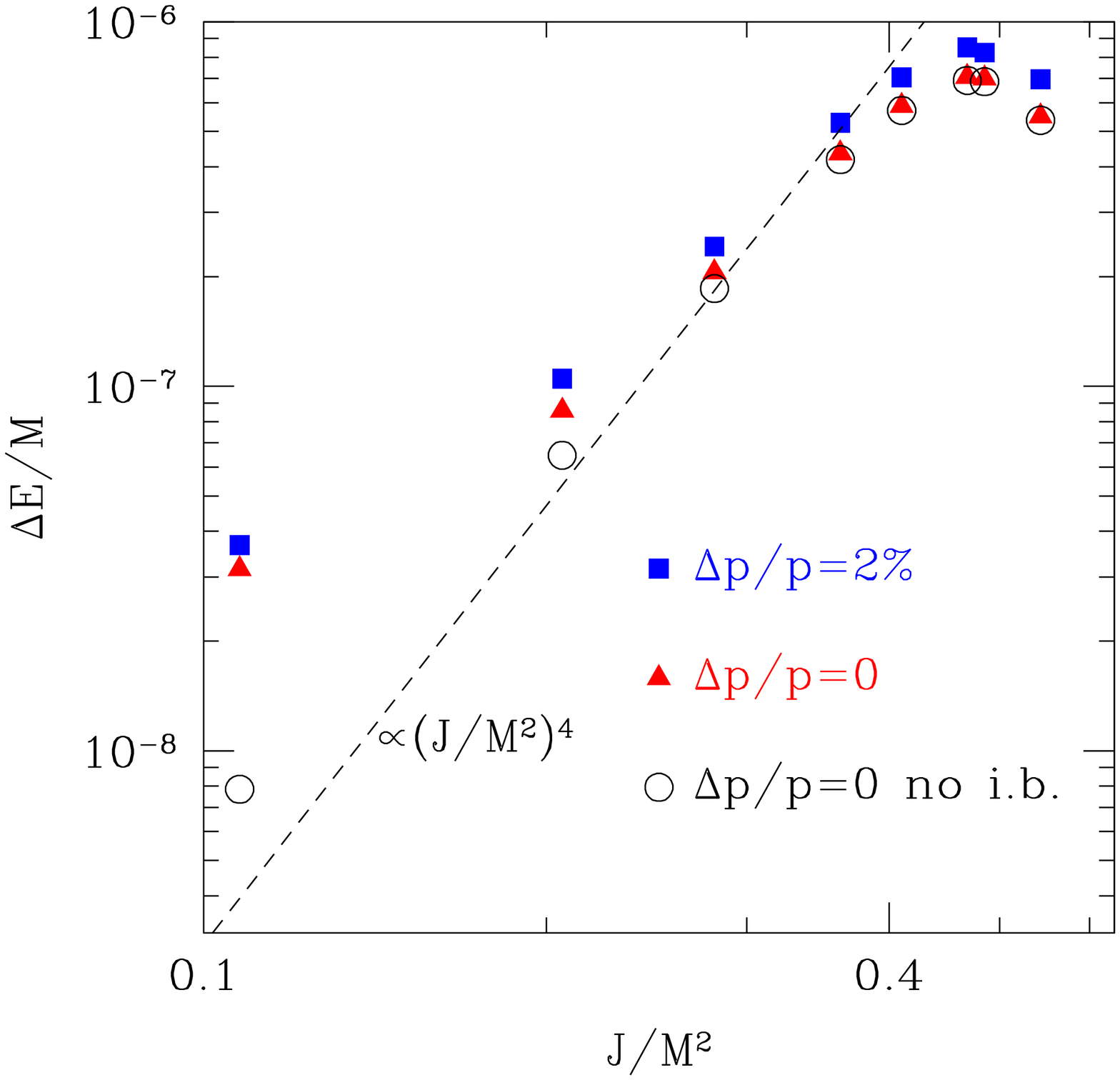}
\hskip 0.5cm
\centering \includegraphics[angle=0,width=6.2cm]{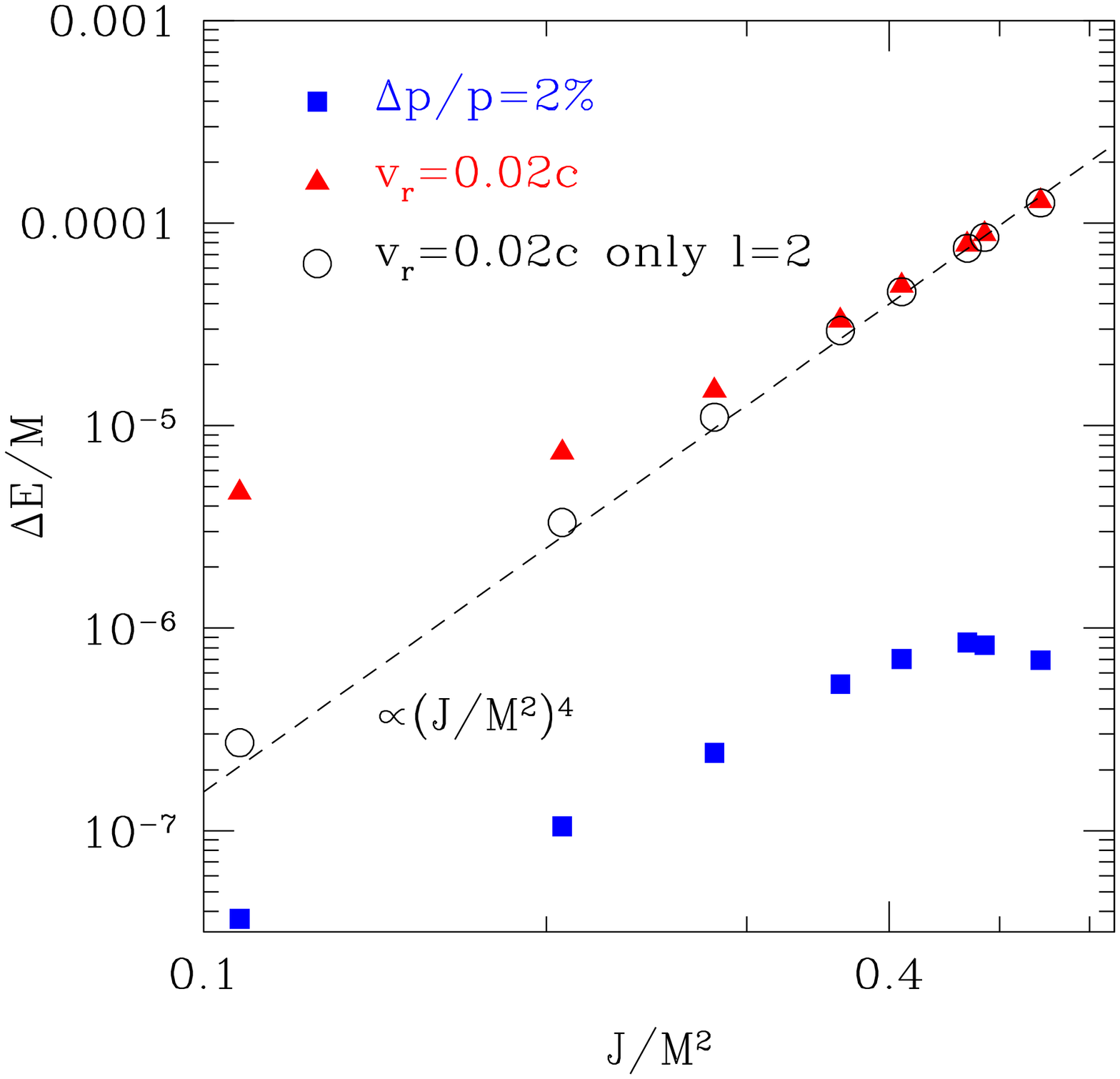}
\caption{Energy carried by the emitted gravitational waves during the
  collapse for different values of the rotation parameter $J/M^2$ and
  initial perturbations. \textit{Left panel:} Filled squares and
  triangles refer respectively to models with a 2\% pressure
  perturbation and to models that are unperturbed. Open triangles, on
  the other hand, are the same as the filled ones but exclude the
  initial burst in the waveforms (see Sect.~\ref{ropg}).
  \textit{Right panel:} Filled triangles refer to models perturbed
  with an inward uniform radial velocity of 0.02 and the open circles
  to the same models but considering only the $\ell=2$ contribution to
  the energy; filled squares refer again to pressure-perturbed models
  and are used as a reference. In both panels the measurements are
  made at a coordinate distance of $50~M$ and the dashed lines
  indicate a scaling $\propto (J/M^2)^4$.}
\label{fig:energy} 
\end{figure}

In Fig.~\ref{fig:energy} we present a summary of the efficiency in the
collapse to black hole by reporting in a log-log plot the emitted
energy as a function of the initial stellar rotation rate parameter
$J/M^2$ and for different initial perturbations. A discussion on how
to calculate this energy from the gauge-invariant quantities can be
found in~\cite{Nagar05,Baiotti06c} and it has been here calculated for
an observer at a coordinate distance of $50~M$. The left panel of
Fig.~\ref{fig:energy}, in particular, highlights the influence of
pressure perturbations and shows with filled squares and triangles
models with a 2\% pressure perturbation and unperturbed models,
respectively. Open circles, on the other hand, are the same as the
filled ones but when the initial burst in the waveforms is excluded
(see Sect.~\ref{ropg} for a discussion).

Clearly, and as first pointed out in~\cite{Stark85}, the efficiency
follows a behaviour of the type $\Delta E/E \propto (J/M^2)^4$ almost
up to the largest rotations rates that yield equilibrium models in
uniform rotation, i.e.  $J/M^2 \lesssim 0.54$. After that, the
efficiency does not grow further and this represents a difference with
respect to what found in~\cite{Stark85}, where the efficiency essentially
saturated at very large rotation rates (We recall that the rather
crude way of introducing rotation in the initial models allowed to
reach values as large as $J/M^2\simeq 0.9$ in \cite{Stark85}.). As
mentioned in Sect.~\ref{doc}, this is probably due to the increased
centrifugal support that these models experience and that effectively
slows down the growth-time for the dynamical instability (cf. right
panel of Fig.~\ref{fig:rhomax_evol}). The value of $J/M^2$ at which
the maximum efficiency is reached depends on the rapidity of the
collapse and hence on the initial perturbation. For models with a
small or zero initial perturbation, the maximum is located at
$J/M^2\simeq 0.4$, while for more rapid collapses (as those shown in
the right panel of Fig.~\ref{fig:energy}), this happens at higher
rotation rates. Note also that the efficiency does not follow a
power-law behaviour at very small values of $J/M^2$. A comparison with
the efficiency calculated not including the initial burst (open
circles) shows that this is just the result of the initial spurious
gravitational wave signal that, as mentioned above, can represent a
significant fraction of the whole signal at low rotation rates.

The right panel of Fig.~\ref{fig:energy}, on the other hand,
highlights the influence on the energy-efficiency of velocity
perturbations, with the filled triangle referring to models perturbed
with an inward uniform radial velocity of 0.02 and with the open
circles referring to the same models but when considering only the
$\ell=2$ contribution to the energy. Clearly, velocity perturbations
do not alter the overall scaling with rotation but do produce a
significant increase in the efficiency, which can easily become two
orders of magnitude larger than the one produced with pressure
perturbations (this is shown with filled squares as a reference). This
enhanced emission is essentially the result of a more rapid change in
mass quadrupole (indeed the amplitude of the $\ell=2$ mode is always
larger than the corresponding mode in the cases of pressure depletion)
but it also receives a contribution from higher-order multipoles,
especially from the $\ell=4$ and at low-rotation rates (cf. filled
triangles and open circles in the right panel of
Fig.~\ref{fig:energy}). The evidence that the $\ell=4$ contribution to
the overall energy is rather similar at all rotation rates seems to
indicate that this is just an artifact of the initial perturbation and
that a very clear scaling $\propto (J/M^2)^4$ is recovered when
considering the $\ell=2$ contribution only (open circles). This
result, on the other hand, also highlights that the study of the
multipolar structure of the gravitational-wave emission from the
collapse can be used to deduce the dynamical and kinematical
properties of the star at the time of the collapse.

The gravitational-wave information computed here can also be used to
determine the detectability of these sources~\cite{Nagar05,Baiotti06c} so
that and in the case of an interferometric detector with the sensitivity
of Virgo and the signal coming from the gravitational-collapse only, we
set an upper limit for the characteristic amplitude produced in the
collapse of a rapidly and uniformly-rotating polytropic star at 10 kpc to
be $h_c = 5.77 \times 10^{-22}(M/M_{\odot})$ at a characteristic
frequency {$f_c=931$ Hz.}  In the case of a detector with the sensitivity
of LIGO I, instead, we obtain $h_c = 5.46 \times 10^{-22}(M/M_{\odot})$
at {$f_c=531$ Hz}.  The resulting signal-to-noise ratios are then
$(S/N)_{_{\rm D1-D4}}^{\rm Virgo} \simeq 0.27-2.1$, $(S/N)_{_{\rm
D1-D4}}^{^{\rm advLIGO}} \simeq 1.2-11$, and $(S/N)_{_{\rm D1-D4}}^{^{\rm
Dual}} \simeq 3.3-28$ for detectors such as Virgo/LIGO, advanced LIGO or
Dual~\cite{Bonaldi:Dual2003}.



\section{Conclusions}
\label{sec:conclusion}

We have provided details and presented additional results on the
numerical study of the gravitational-wave emission from the collapse of
neutron stars to rotating black holes in three
dimensions~\cite{Baiotti04b,Baiotti06}. In particular, we have discussed
the advantages and disadvantages of the use of the excision technique and
how alternative approaches to that of excision can used with great
success to extract the \textit{complete} gravitational-wave
signal~\cite{Baiotti06}. 

As a first step towards the characterization of these sources of
gravitational waves, we have presented a systematic discussion of the
influence that rotation and different perturbations have on the waveforms
and hence on the energy emitted in gravitational waves. In particular, a
systematic analysis of the waveforms calculated under different initial
rotation rates has provided the first estimates in full General
Relativity and for rapidly rotating stars of the growth-time for the
dynamical instability to axisymmetric perturbations and confirmed the
existence of a precise power-law scaling of the energy-efficiency in
terms of rotation parameter $J/M^2$.

We have also shown that the pressure perturbations traditionally used to
trigger the collapse do not affect sensitively the dynamics of the matter
and of the trapped surfaces as long as they are very small. Excessively
large pressure depletions, on the other hand, can change significantly
the way the collapse proceeds as well as artificially amplify the
energy-efficiency in the emission of gravitational waves. This clarifies
the source of the differences between our estimates for the efficiency
and those made in axisymmetry in~\cite{Stark85}.  Furthermore, the study
of the waveforms produced with perturbations of different amplitude and
type has also made it possible to isolate the part of the signal produced
by the actual collapse from the spurious one which should instead be
related to initial violations of the constraint equations and which is
produced either from the interpolation of the initial data onto a
Cartesian grid or from the introduction of the initial
perturbations. While it is still unclear whether this initial signal
reflects a consistent response of the star to a perturbation and can
therefore be associated to a $w$ mode, a number of considerations seem to
support this hypothesis.

Overall, the found results indicate that the gravitational collapse of
axisymmetric neutrons stars to rotating black holes is not an efficient
process for converting the binding energy into gravitational waves, with
an overall efficiency $\delta M/M \simeq 10^{-7}-10^{-6}$ for uniformly
rotating models. This efficiency, however, can be increased of up to two
orders of magnitude if velocity perturbations are present in the
collapsing star and it is possible that similar conclusions may be valid
also for the collapse of differentially rotating models.

As a concluding remark we note that while this work, together with the
ones preceding it~\cite{Baiotti04,Baiotti04b,Baiotti06}, has provided a
full and consistent picture of the gravitational-wave emission from the
collapse of neutron stars to rotating black holes, it represents only a
very idealized description of this process. Additional and considerable
work is still needed both in the modelling of the matter (through
improved equations of state, the inclusion of the contributions coming
from magnetic fields, radiation transport, multifluids, a solid crust,
etc.) and in the numerical techniques needed to handle this improved
modelling. Both aspects will represent the focus of our future research.

\hskip 2.0cm

\ack 

It is a pleasure to thank Erik Schnetter for his help with the mesh
refinement and for many useful discussions. The computations presented
here were performed on the {\it Albert} cluster at the University of
Parma and the {\em Peyote} cluster at the Albert-Einstein-Institut.

\hskip 2.0cm

\bibliographystyle{iopart-num}

\bibliography{references}


\end{document}